\documentclass[12pt]{article}
\usepackage{amssymb}
\begin{document}
\newcommand{\beq}{\begin{equation}}
\newcommand{\eeq}{\end{equation}}
\newcommand{\beqa}{\begin{eqnarray}}
\newcommand{\eeqa}{\end{eqnarray}}
\newcommand{\beqar}{\begin{eqnarray*}}
\newcommand{\eeqar}{\end{eqnarray*}}
\newcommand{\al}{\alpha}
\newcommand{\be}{\beta}
\newcommand{\del}{\delta}
\newcommand{\D}{\Delta}
\newcommand{\eps}{\epsilon}
\newcommand{\ga}{\gamma}
\newcommand{\Ga}{\Gamma}
\newcommand{\ka}{\kappa}
\newcommand{\nn}{\nonumber}
\newcommand{\inn}{\!\cdot\!}
\newcommand{\h}{\eta}
\newcommand{\ii}{\iota}
\newcommand{\kk}{\varphi}
\newcommand\F{{}_3F_2}
\newcommand{\la}{\lambda}
\newcommand{\La}{\Lambda}
\newcommand{\na}{\prt}
\newcommand{\Om}{\Omega}
\newcommand{\om}{\omega}
\newcommand{\p}{\phi}
\newcommand{\sig}{\sigma}
\renewcommand{\t}{\theta}
\newcommand{\z}{\zeta}
\newcommand{\ssc}{\scriptscriptstyle}
\newcommand{\eg}{{\it e.g.,}\ }
\newcommand{\ie}{{\it i.e.,}\ }
\newcommand{\labell}[1]{\label{#1}} %{\label{#1}} %
\newcommand{\reef}[1]{(\ref{#1})}
\newcommand\prt{\partial}
\newcommand\veps{\varepsilon}
\newcommand{\pol}{\varepsilon}
\newcommand\vp{\varphi}
\newcommand\ls{\ell_s}
\newcommand\dS{\dot{\cal S}}
\newcommand\dB{\dot{B}}
\newcommand\dG{\dot{G}}
\newcommand\ddG{\dot{\dot{G}}}
\newcommand\dP{\dot{\phi}}
\newcommand\cF{{\cal F}}
\newcommand\cA{{\cal A}}
\newcommand\cS{{\cal S}}
\newcommand\cT{{\cal T}}
\newcommand\cV{{\cal V}}
\newcommand\cL{{\cal L}}
\newcommand\cM{{\cal M}}
\newcommand\cN{{\cal N}}
\newcommand\cG{{\cal G}}
\newcommand\cH{{\cal H}}
\newcommand\cI{{\cal I}}
\newcommand\cJ{{\cal J}}
\newcommand\cl{{\iota}}
\newcommand\cP{{\cal P}}
\newcommand\cQ{{\cal Q}}
\newcommand\cg{{\it g}}
\newcommand\cR{{\cal R}}
\newcommand\cB{{\cal B}}
\newcommand\cO{{\cal O}}
\newcommand\tcO{{\tilde {{\cal O}}}}
\newcommand\bg{\bar{g}}
\newcommand\bb{\bar{b}}
\newcommand\bH{\bar{H}}
\newcommand\bX{\bar{X}}
\newcommand\bK{\bar{K}}
\newcommand\bA{\bar{A}}
\newcommand\bZ{\bar{Z}}
\newcommand\bxi{\bar{\xi}}
\newcommand\bphi{\bar{\phi}}
\newcommand\bpsi{\bar{\psi}}
\newcommand\bprt{\bar{\prt}}
\newcommand\bet{\bar{\eta}}
\newcommand\btau{\bar{\tau}}
\newcommand\bnabla{\bar{\nabla}}
\newcommand\hF{\hat{F}}
\newcommand\hA{\hat{A}}
\newcommand\hT{\hat{T}}
\newcommand\htau{\hat{\tau}}
\newcommand\hD{\hat{D}}
\newcommand\hf{\hat{f}}
\newcommand\hg{\hat{g}}
\newcommand\hp{\hat{\phi}}
\newcommand\hi{\hat{i}}
\newcommand\ha{\hat{a}}
\newcommand\hb{\hat{b}}
\newcommand\hQ{\hat{Q}}
\newcommand\hP{\hat{\Phi}}
\newcommand\hS{\hat{S}}
\newcommand\hX{\hat{X}}
\newcommand\tL{\tilde{\cal L}}
\newcommand\hL{\hat{\cal L}}
\newcommand\tG{{\widetilde G}}
\newcommand\tg{{\widetilde g}}
\newcommand\tphi{{\widetilde \phi}}
\newcommand\tPhi{{\widetilde \Phi}}
\newcommand\td{{\tilde d}}
\newcommand\tk{{\tilde k}}
\newcommand\tf{{\tilde f}}
\newcommand\ta{{\tilde a}}
\newcommand\tb{{\tilde b}}
\newcommand\tc{{\tilde c}}
\newcommand\tR{{\tilde R}}
\newcommand\teta{{\tilde \eta}}
\newcommand\tF{{\widetilde F}}
\newcommand\tK{{\widetilde K}}
\newcommand\tE{{\widetilde E}}
\newcommand\tpsi{{\tilde \psi}}
\newcommand\tX{{\widetilde X}}
\newcommand\tD{{\widetilde D}}
\newcommand\tO{{\widetilde O}}
\newcommand\tS{{\tilde S}}
\newcommand\tB{{\widetilde B}}
\newcommand\tA{{\widetilde A}}
\newcommand\tT{{\widetilde T}}
\newcommand\tC{{\widetilde C}}
\newcommand\tV{{\widetilde V}}
\newcommand\thF{{\widetilde {\hat {F}}}}
\newcommand\Tr{{\rm Tr}}
\newcommand\tr{{\rm tr}}
\newcommand\STr{{\rm STr}}
\newcommand\hR{\hat{R}}
\newcommand\M[2]{M^{#1}{}_{#2}}

\newcommand\bS{\textbf{ S}}
\newcommand\bI{\textbf{ I}}
\newcommand\bJ{\textbf{ J}}

\newcommand\MZ{\mathbb{Z}}
\newcommand\MR{\mathbb{R}}

%\begin{document}
\begin{titlepage}
\begin{center}

\vskip 2 cm
{\LARGE \bf More on closed string effective actions \\ \vskip 0.75  cm  at order $\alpha'^2$}\\
\vskip 1.25 cm
  Hadi Gholian\footnote{hadi.gholianaval@mail.um.ac.ir} and  Mohammad R. Garousi\footnote{garousi@um.ac.ir}

\vskip 1 cm
{{\it Department of Physics, Faculty of Science, Ferdowsi University of Mashhad\\}{\it P.O. Box 1436, Mashhad, Iran}\\}
\vskip .1 cm
 \end{center}

\begin{abstract}
Recent progress in string theory has unveiled the discovery of NS-NS couplings in bosonic and heterotic effective actions at order $\alpha'^2$, which were achieved by imposing $O(1,1)$ symmetry on the circle reduction of classical effective actions. While the bosonic theory features 25 couplings, the heterotic theory encompasses 24 parity-even and 3 parity-odd couplings, excluding the pure gravity couplings. In this study, we redefine the even-parity couplings in the bosonic and heterotic theories through the application of appropriate field redefinitions, resulting in 10 and 8 couplings, respectively. To establish the validity of these couplings, a cosmological reduction is conducted, demonstrating that the cosmological couplings in the heterotic theory vanish, subject to one-dimensional field redefinitions that include the lapse function and total derivative terms. Additionally, it is observed that the cosmological couplings in the bosonic theory can be expressed as $\mathrm{tr}(\dS^6)$. These results are consistent with existing literature, where such behavior is attributed to the pure gravity component of the couplings. Furthermore, the consistency of the obtained couplings with 4-point string theory S-matrix elements is confirmed.
\end{abstract}
\end{titlepage}

\section{Introduction}

The spectrum of quantized free closed strings encompasses both massless states and an infinite tower of massive states. Among these massless states, a notable one is the spin-2 state, which showcases the potential of string theory as a promising candidate for a consistent theory of quantum gravity.
To delve into physics within this framework, it is convenient to employ an effective action that incorporates only the massless fields. The influence of the massive fields is revealed through higher derivatives of the massless fields, commonly known as $\alpha'$-corrections, which encompass classical and loop corrections. Determining these actions can be achieved through various approaches, such as the S-matrix method \cite{Gross:1986iv, Gross:1986mw}, the sigma-mode method \cite{Grisaru:1986vi, Freeman:1986zh}, or by exploring different symmetries in string theory.
In the past, the imposition of local supersymmetry on the effective action has been utilized to derive leading-order classical effective actions in superstring theories (see, for example,  \cite{Becker:2007zj}). This local symmetry, which necessitates the consideration of both bosonic and fermionic fields within the theory, can also be employed to study the $\alpha'$ corrections \cite{Gates:1986dm, Gates:1985wh,Bergshoeff:1986wc}.

Another intriguing symmetry in perturbative bosonic string theory or superstring theory is T-duality \cite{Giveon:1994fu,Alvarez:1994dn}, which arises when the theory is compactified on a torus. When integrating out the massive modes, T-duality emerges as a symmetry in the effective actions. It has been demonstrated in \cite{Sen:1991zi,Hohm:2014sxa} that the dimensional reduction of the classical effective actions of bosonic string theory and heterotic string theory at each order of $\alpha'$ remains invariant under $O(d,d,\MR)$ transformations.
By assuming that the classical effective actions of string theory are background independent \cite{Garousi:2022ovo}, it becomes possible to explore a specific closed background featuring a circular dimension. Applying the non-geometrical $O(1,1,\MZ)$ symmetry to the circular reduction of the independent covariant and gauge invariant couplings in closed spacetime manifolds allows us to determine the coefficients of background-independent couplings for closed spacetime manifolds. This technique has proven successful in determining the NS-NS couplings in closed spacetime manifolds up to order $\alpha'^3$ \cite{Garousi:2019wgz,Garousi:2019mca,Garousi:2023kxw,Garousi:2020gio,Garousi:2020lof}.

At the leading order of $\alpha'$, T-duality reproduces the standard effective action, given by
\beqa
\bS^{(0)}
=-\frac{2}{\kappa^2} \int d^{D}x \, e^{-2\phi}\sqrt{-G} \left(  R + 4\nabla_{\mu}\phi \nabla^{\mu}\phi-\frac{1}{12}H^2\right)\,.\labell{action1}
\eeqa
At higher orders of $\alpha'$, and for closed spacetime manifolds that have no boundary where data would be present, there is freedom in utilizing the most general higher-derivative field redefinitions \cite{Gross:1986iv,Tseytlin:1986ti,Deser:1986xr,Garousi:2019cdn}. If this freedom is employed to construct the 8 independent parity-even covariant couplings at order $\alpha'$, T-duality determines their background-independent couplings up to one overall factor, as demonstrated in \cite{Garousi:2019wgz},
  \beqa
 \bS^{(1)}_{MT}&=&\frac{-2\alpha'a_1 }{\kappa^2}\int d^{D}x\, e^{-2\phi}\sqrt{-G}\Big(   R_{\alpha \beta \gamma \delta} R^{\alpha \beta \gamma \delta} -\frac{1}{2}H_{\alpha}{}^{\delta \epsilon} H^{\alpha \beta \gamma} R_{\beta  \gamma \delta\epsilon}\nn\\
&&\qquad\qquad\qquad\qquad\qquad\quad+\frac{1}{24}H_{\epsilon\delta \zeta}H^{\epsilon}{}_{\alpha}{}^{\beta}H^{\delta}{}_{\beta}{}^{\gamma}H^{\zeta}{}_{\gamma}{}^{\alpha}-\frac{1}{8}H_{\alpha \beta}{}^{\delta} H^{\alpha \beta \gamma} H_{\gamma}{}^{\epsilon \zeta} H_{\delta \epsilon \zeta}\Big)\,.\labell{S1bf}
\eeqa
The above action is the Metsaev-Tseytlin action \cite{Metsaev:1987zx}, derived from the S-matrix elements. For the bosonic string theory, $a_1=1/4$, and for the heterotic theory, it is $a_1=1/8$. 

There are various other forms for the effective action at this order that are related to the above action through field redefinitions, such as the action in the Meissner scheme \cite{Meissner:1996sa}, which is given by
\beqa
\bS^{(1)}_M&\!\!\!\!\!=\!\!\!\!\!&-\frac{2\alpha' a_1}{\kappa^2}\int d^{D} x\sqrt{-G} e^{-2\Phi}\Big[R_{GB}^2+\frac{1}{24} H_{\alpha }{}^{\delta \epsilon } H^{\alpha \beta
\gamma } H_{\beta \delta }{}^{\varepsilon } H_{\gamma \epsilon
\varepsilon } -  \frac{1}{8} H_{\alpha \beta }{}^{\delta }
H^{\alpha \beta \gamma } H_{\gamma }{}^{\epsilon \varepsilon }
H_{\delta \epsilon \varepsilon }\nn\\&&  + \frac{1}{144} H_{\alpha
\beta \gamma } H^{\alpha \beta \gamma } H_{\delta \epsilon
\varepsilon } H^{\delta \epsilon \varepsilon }+ H_{\alpha }{}^{
\gamma \delta } H_{\beta \gamma \delta } R^{\alpha
\beta } -  \frac{1}{6} H_{\alpha \beta \gamma } H^{\alpha \beta
\gamma } R  -
\frac{1}{2} H_{\alpha }{}^{\delta \epsilon } H^{\alpha \beta
\gamma } R_{\beta \gamma \delta \epsilon }\nn\\&& -
\frac{2}{3} H_{\beta \gamma \delta } H^{\beta \gamma \delta }
\nabla_{\alpha }\nabla^{\alpha }\Phi + \frac{2}{3} H_{\beta
\gamma \delta } H^{\beta \gamma \delta } \nabla_{\alpha }\Phi
\nabla^{\alpha }\Phi + 8 R \nabla_{\alpha }\Phi
\nabla^{\alpha }\Phi - 16
R_{\alpha \beta } \nabla^{\alpha }\Phi \nabla^{\beta
}\Phi \nn\\&&+ 16 \nabla_{\alpha }\Phi \nabla^{\alpha
}\Phi \nabla_{\beta }\nabla^{\beta }\Phi - 16 \nabla_{\alpha }\Phi \nabla^{\alpha }\Phi \nabla_{
\beta }\Phi \nabla^{\beta }\Phi + 2 H_{\alpha }{}^{\gamma
\delta } H_{\beta \gamma \delta } \nabla^{\beta
}\nabla^{\alpha }\Phi \Big]\,,\labell{Mis}
\eeqa
 where  $R_{GB}^2= R_{\alpha \beta \gamma \delta} R^{\alpha \beta \gamma \delta}-4R^{\alpha\beta}R_{\alpha\be}+R^2$ represents the Gauss-Bonnet couplings.

Using the most general higher-derivative field redefinition freedom, it is found that at order $\alpha'$, there are 60 independent parity-even covariant couplings in closed spacetime manifolds \cite{Garousi:2019cdn}. To determine their background-independent couplings through T-duality, the effective action at order $\alpha'$ is required, as it demonstrates the observation that the form of the effective action at order $\alpha'^2$ depends on the form of the effective action at order $\alpha'$ \cite{Bento:1990nv}. Choosing \reef{S1bf} for the effective action at order $\alpha'$, the T-duality fixes the coefficients of the 60 parity-even couplings at order $\alpha'^2$ in bosonic string theory as follows \cite{Garousi:2019mca}\footnote{Note that there is a typo in the overall coefficient of $\bS^{(2)B}_{MT}$ in \cite{Garousi:2019mca}, i.e., the factor $a_1$ in \cite{Garousi:2019mca} should be $a_1^2$.}:
\beqa
&&\bS^{(2)B}_{MT}=\frac{-2\alpha'^2a_1^2 }{\kappa^2}\int d^{26}x \,e^{-2\Phi}\sqrt{-G}\Big[  - \frac{1}{12} H_{\alpha}{}^{\delta \epsilon} H^{\alpha \beta 
\gamma} H_{\beta \delta}{}^{\zeta} H_{\gamma}{}^{\iota \kappa} 
H_{\epsilon \iota}{}^{\mu} H_{\zeta \kappa \mu}\labell{S2f}\\&& + 
\frac{1}{30} H_{\alpha \beta}{}^{\delta} H^{\alpha \beta 
\gamma} H_{\gamma}{}^{\epsilon \zeta} H_{\delta}{}^{\iota 
\kappa} H_{\epsilon \zeta}{}^{\mu} H_{\iota \kappa \mu} + 
\frac{3}{10} H_{\alpha \beta}{}^{\delta} H^{\alpha \beta 
\gamma} H_{\gamma}{}^{\epsilon \zeta} H_{\delta 
\epsilon}{}^{\iota} H_{\zeta}{}^{\kappa \mu} H_{\iota \kappa 
\mu} \nn\\&&+ \frac{13}{20} H_{\alpha}{}^{\epsilon \zeta} 
H_{\beta}{}^{\iota \kappa} H_{\gamma \epsilon \zeta} H_{\delta 
\iota \kappa} R^{\alpha \beta \gamma \delta} + \frac{2}{5} 
H_{\alpha}{}^{\epsilon \zeta} H_{\beta \epsilon}{}^{\iota} 
H_{\gamma \zeta}{}^{\kappa} H_{\delta \iota \kappa} R^{\alpha 
\beta \gamma \delta} + \frac{18}{5} H_{\alpha 
\gamma}{}^{\epsilon} H_{\beta}{}^{\zeta \iota} H_{\delta 
\zeta}{}^{\kappa} H_{\epsilon \iota \kappa} R^{\alpha \beta 
\gamma \delta} \nn\\&&-  \frac{43}{5} H_{\alpha \gamma}{}^{\epsilon} 
H_{\beta}{}^{\zeta \iota} H_{\delta \epsilon}{}^{\kappa} 
H_{\zeta \iota \kappa} R^{\alpha \beta \gamma \delta} -  
\frac{16}{5} H_{\alpha \gamma}{}^{\epsilon} H_{\beta 
\delta}{}^{\zeta} H_{\epsilon}{}^{\iota \kappa} H_{\zeta \iota 
\kappa} R^{\alpha \beta \gamma \delta} - 2 H_{\beta 
\epsilon}{}^{\iota} H_{\delta \zeta \iota} 
R_{\alpha}{}^{\epsilon}{}_{\gamma}{}^{\zeta} R^{\alpha \beta 
\gamma \delta} \nn\\&&- 2 H_{\beta \delta}{}^{\iota} H_{\epsilon 
\zeta \iota} R_{\alpha}{}^{\epsilon}{}_{\gamma}{}^{\zeta} 
R^{\alpha \beta \gamma \delta} -  \frac{4}{3} 
R_{\alpha}{}^{\epsilon}{}_{\gamma}{}^{\zeta} R^{\alpha \beta 
\gamma \delta} R_{\beta \zeta \delta \epsilon} + \frac{4}{3} 
R_{\alpha \beta}{}^{\epsilon \zeta} R^{\alpha \beta \gamma 
\delta} R_{\gamma \epsilon \delta \zeta} + 3 
H_{\beta}{}^{\zeta \iota} H_{\epsilon \zeta \iota} R^{\alpha 
\beta \gamma \delta} R_{\gamma}{}^{\epsilon}{}_{\alpha \delta} 
\nn\\&&+ 2 H_{\beta \epsilon}{}^{\iota} H_{\delta \zeta \iota} 
R^{\alpha \beta \gamma \delta} 
R_{\gamma}{}^{\epsilon}{}_{\alpha}{}^{\zeta} + 2 H_{\alpha 
\beta \epsilon} H_{\delta \zeta \iota} R^{\alpha \beta \gamma 
\delta} R_{\gamma}{}^{\epsilon \zeta \iota} + \frac{13}{10} 
H_{\alpha}{}^{\gamma \delta} H_{\beta \gamma}{}^{\epsilon} 
H_{\delta}{}^{\zeta \iota} H_{\epsilon \zeta \iota} 
\nabla^{\beta}\nabla^{\alpha}\Phi\nn\\&& + \frac{13}{5} 
H_{\gamma}{}^{\epsilon \zeta} H_{\delta \epsilon \zeta} 
R_{\alpha}{}^{\gamma}{}_{\beta}{}^{\delta} 
\nabla^{\beta}\nabla^{\alpha}\Phi -  \frac{52}{5} H_{\beta 
\delta}{}^{\zeta} H_{\gamma \epsilon \zeta} 
R_{\alpha}{}^{\gamma \delta \epsilon} 
\nabla^{\beta}\nabla^{\alpha}\Phi -  \frac{26}{5} H_{\alpha 
\gamma \epsilon} H_{\beta \delta \zeta} R^{\gamma \delta 
\epsilon \zeta} \nabla^{\beta}\nabla^{\alpha}\Phi \nn\\&&+ 
\frac{13}{5} \nabla^{\beta}\nabla^{\alpha}\Phi 
\nabla_{\epsilon}H_{\beta \gamma \delta} 
\nabla^{\epsilon}H_{\alpha}{}^{\gamma \delta}+ \frac{13}{10} 
H_{\beta \gamma}{}^{\epsilon} H^{\beta \gamma \delta} 
H_{\delta}{}^{\zeta \iota} \nabla^{\alpha}\Phi 
\nabla_{\iota}H_{\alpha \epsilon \zeta}+ \frac{1}{20} 
H_{\alpha}{}^{\delta \epsilon} H^{\alpha \beta \gamma} 
\nabla_{\iota}H_{\delta \epsilon \zeta} 
\nabla^{\iota}H_{\beta \gamma}{}^{\zeta} \nn\\&& -  \frac{13}{20} 
H_{\alpha}{}^{\beta \gamma} H_{\delta \epsilon}{}^{\iota} 
H^{\delta \epsilon \zeta} \nabla^{\alpha}\Phi 
\nabla_{\iota}H_{\beta \gamma \zeta}  + \frac{1}{5} 
H_{\alpha}{}^{\delta \epsilon} H^{\alpha \beta \gamma} 
\nabla_{\zeta}H_{\gamma \epsilon \iota} 
\nabla^{\iota}H_{\beta \delta}{}^{\zeta}\nn\\&& -  \frac{6}{5} 
H_{\alpha}{}^{\delta \epsilon} H^{\alpha \beta \gamma} 
\nabla_{\iota}H_{\gamma \epsilon \zeta} 
\nabla^{\iota}H_{\beta \delta}{}^{\zeta} -  \frac{6}{5} 
H_{\alpha \beta}{}^{\delta} H^{\alpha \beta \gamma} 
\nabla_{\zeta}H_{\delta \epsilon \iota} 
\nabla^{\iota}H_{\gamma}{}^{\epsilon \zeta} + \frac{17}{10} 
H_{\alpha \beta}{}^{\delta} H^{\alpha \beta \gamma} 
\nabla_{\iota}H_{\delta \epsilon \zeta} 
\nabla^{\iota}H_{\gamma}{}^{\epsilon \zeta}\Big]\,,\nn
\eeqa
where $a_1=1/4$. To compare the above couplings with the S-matrix elements, it is appropriate to change the order $\alpha'$ action \reef{S1bf} to the Meissner scheme \reef{Mis}, where the graviton propagator does not receive $\alpha'$ corrections. By using field redefinitions at order $\alpha'$ that transform the action \reef{S1bf} to the action \reef{Mis}, we have discovered that the same field redefinitions also transform the action \reef{S2f} to the following action:
\beqa
&&\bS^{(2)B}_{M}=\frac{2\alpha'^2 a_1^2 }{\kappa^2}\int d^{26} x\sqrt{-G} e^{-2\Phi}\Big[\frac{1}{12} H_{\alpha }{}^{\delta \epsilon } H^{\alpha \beta 
\gamma } H_{\beta \delta }{}^{\zeta } H_{\gamma }{}^{\iota 
\kappa } H_{\epsilon \iota }{}^{\mu } H_{\zeta \kappa \mu }\labell{Sb}\\&& -  
\frac{1}{30} H_{\alpha \beta }{}^{\delta } H^{\alpha \beta 
\gamma } H_{\gamma }{}^{\epsilon \zeta } H_{\delta }{}^{\iota 
\kappa } H_{\epsilon \zeta }{}^{\mu } H_{\iota \kappa \mu } -  
\frac{1}{20} H_{\alpha \beta }{}^{\delta } H^{\alpha \beta 
\gamma } H_{\gamma }{}^{\epsilon \zeta } H_{\delta \epsilon 
}{}^{\iota } H_{\zeta }{}^{\kappa \mu } H_{\iota \kappa \mu } \nn\\&&
+ \frac{4}{3} R_{\alpha }{}^{\epsilon }{}_{\gamma }{}^{
\zeta } R^{\alpha \beta \gamma \delta } 
R_{\beta \zeta \delta \epsilon } -  \frac{4}{3} 
R_{\alpha \beta }{}^{\epsilon \zeta } 
R^{\alpha \beta \gamma \delta } R_{\gamma 
\epsilon \delta \zeta } -  \frac{2}{5} H_{\alpha }{}^{\delta 
\epsilon } H^{\alpha \beta \gamma } H_{\beta }{}^{\zeta \iota 
} H_{\delta \zeta }{}^{\kappa } R_{\gamma \epsilon 
\iota \kappa } \nn\\&&+ 2 H_{\alpha }{}^{\delta \epsilon } H^{\alpha 
\beta \gamma } R_{\beta }{}^{\zeta }{}_{\delta 
}{}^{\iota } R_{\gamma \zeta \epsilon \iota } -  
\frac{3}{20} H_{\alpha \beta }{}^{\delta } H^{\alpha \beta 
\gamma } H_{\epsilon \zeta }{}^{\kappa } H^{\epsilon \zeta 
\iota } R_{\gamma \iota \delta \kappa } - 2 H^{\alpha 
\beta \gamma } H^{\delta \epsilon \zeta } R_{\alpha 
\beta \delta }{}^{\iota } R_{\gamma \iota \epsilon 
\zeta }\nn\\&& - 2 H_{\alpha }{}^{\delta \epsilon } H^{\alpha \beta 
\gamma } R_{\beta }{}^{\zeta }{}_{\delta }{}^{\iota } 
R_{\gamma \iota \epsilon \zeta } + 2 H_{\alpha 
}{}^{\delta \epsilon } H^{\alpha \beta \gamma } 
R_{\beta }{}^{\zeta }{}_{\gamma }{}^{\iota } 
R_{\delta \zeta \epsilon \iota } + H_{\alpha \beta 
}{}^{\delta } H^{\alpha \beta \gamma } R_{\gamma 
}{}^{\epsilon \zeta \iota } R_{\delta \zeta \epsilon 
\iota }\nn\\&& -  \frac{3}{5} H_{\alpha \beta }{}^{\delta } H^{\alpha 
\beta \gamma } H_{\gamma }{}^{\epsilon \zeta } H_{\epsilon 
}{}^{\iota \kappa } R_{\delta \iota \zeta \kappa } -  
\frac{8}{5} H_{\alpha }{}^{\delta \epsilon } H^{\alpha \beta 
\gamma } H_{\beta \delta }{}^{\zeta } H_{\gamma }{}^{\iota 
\kappa } R_{\epsilon \iota \zeta \kappa } + 
\frac{1}{5} H_{\alpha \beta }{}^{\delta } H^{\alpha \beta 
\gamma } H_{\gamma }{}^{\epsilon \zeta } H_{\delta }{}^{\iota 
\kappa } R_{\epsilon \iota \zeta \kappa }\nn\\&& -  
\frac{3}{10} H_{\alpha }{}^{\gamma \delta } H_{\beta \gamma 
}{}^{\epsilon } H_{\delta }{}^{\zeta \iota } H_{\epsilon \zeta 
\iota } \nabla^{\beta }\nabla^{\alpha }\Phi -  \frac{3}{5} 
H_{\gamma \delta }{}^{\zeta } H^{\gamma \delta \epsilon } 
R_{\alpha \epsilon \beta \zeta } \nabla^{\beta 
}\nabla^{\alpha }\Phi -  \frac{12}{5} H_{\alpha }{}^{\gamma 
\delta } H_{\gamma }{}^{\epsilon \zeta } R_{\beta 
\epsilon \delta \zeta } \nabla^{\beta }\nabla^{\alpha }\Phi\nn\\&& + 
\frac{6}{5} H_{\alpha }{}^{\gamma \delta } H_{\beta 
}{}^{\epsilon \zeta } R_{\gamma \epsilon \delta \zeta 
} \nabla^{\beta }\nabla^{\alpha }\Phi -  \frac{3}{5} \nabla^{
\beta }\nabla^{\alpha }\Phi \nabla_{\epsilon }H_{\beta \gamma 
\delta } \nabla^{\epsilon }H_{\alpha }{}^{\gamma \delta } -  
\frac{3}{10} H_{\beta \gamma }{}^{\epsilon } H^{\beta \gamma 
\delta } H_{\delta }{}^{\zeta \iota } \nabla^{\alpha }\Phi 
\nabla_{\iota }H_{\alpha \epsilon \zeta }\nn\\&& + \frac{3}{20} 
H_{\alpha }{}^{\beta \gamma } H_{\delta \epsilon }{}^{\iota } 
H^{\delta \epsilon \zeta } \nabla^{\alpha }\Phi \nabla_{\iota 
}H_{\beta \gamma \zeta } -  \frac{1}{20} H_{\alpha }{}^{\delta 
\epsilon } H^{\alpha \beta \gamma } \nabla_{\iota }H_{\delta 
\epsilon \zeta } \nabla^{\iota }H_{\beta \gamma }{}^{\zeta } - 
 \frac{1}{5} H_{\alpha }{}^{\delta \epsilon } H^{\alpha \beta 
\gamma } \nabla_{\zeta }H_{\gamma \epsilon \iota } 
\nabla^{\iota }H_{\beta \delta }{}^{\zeta } \nn\\&&+ \frac{1}{5} 
H_{\alpha }{}^{\delta \epsilon } H^{\alpha \beta \gamma } 
\nabla_{\iota }H_{\gamma \epsilon \zeta } \nabla^{\iota 
}H_{\beta \delta }{}^{\zeta } + \frac{1}{5} H_{\alpha \beta 
}{}^{\delta } H^{\alpha \beta \gamma } \nabla_{\zeta 
}H_{\delta \epsilon \iota } \nabla^{\iota }H_{\gamma 
}{}^{\epsilon \zeta } -  \frac{1}{5} H_{\alpha \beta 
}{}^{\delta } H^{\alpha \beta \gamma } \nabla_{\iota 
}H_{\delta \epsilon \zeta } \nabla^{\iota }H_{\gamma 
}{}^{\epsilon \zeta }\Big]\,.\nn
\eeqa
It has been shown in \cite{Garousi:2023kxw} that if one chooses \reef{Mis} for the effective action at order $\alpha'$, the T-duality fixes the coefficients of the 60 parity-even couplings at order $\alpha'^2$ in bosonic string theory to exactly match the above couplings.

One can still use field redefinitions that only change the fields at order $\alpha'^2$. Such field redefinitions do not alter the form of the Meissner action at order $\alpha'$, but they do affect the form of the couplings in \reef{Sb}. Since the final form of the couplings and the field redefinitions are not known, we need to consider a specific form for the couplings and check if there are corresponding field redefinitions. Calculations for the couplings at order $\alpha'^3$ have been performed in \cite{Garousi:2020lof}. It has been observed that there are schemes in which the dilaton appears in the couplings only as the overall factor $e^{-2\Phi}$. We also observe that such schemes exist at order $\alpha'^2$. 
To determine the form of the other couplings that involve only the Riemann curvature, $H$, and $\nabla H$, we utilize trial and error to find the minimum number of couplings as in \cite{Garousi:2020lof}. We have discovered that there exist field redefinitions such that the above 27 couplings can be expressed in terms of the following 12 couplings:
\beqa
\bS^{(2)B}_{M}&\!\!\!\!\!\!\!\!=\!\!\!\!\!\!\!\!&-\frac{2\alpha'^2 a_1^2 }{\kappa^2}\int d^{26} x\sqrt{-G} e^{-2\Phi}\Big[-  
\frac{4}{3}R_{\alpha }{}^{\kappa }{}_{\gamma 
}{}^{\lambda }R^{\alpha \beta \gamma \theta } 
R_{\beta \lambda \theta \kappa } + \frac{4}{3} 
R_{\alpha \beta }{}^{\kappa \lambda } 
R^{\alpha \beta \gamma \theta }R_{\gamma 
\kappa \theta \lambda }\nn\\&& - \frac{1}{12} H_{\alpha }{}^{\theta \kappa } H^{\alpha \beta 
\gamma } H_{\beta \theta }{}^{\lambda } H_{\gamma }{}^{\mu \nu 
} H_{\kappa \mu }{}^{\tau } H_{\lambda \nu \tau } + 
\frac{1}{4} H_{\alpha \beta }{}^{\theta } H^{\alpha \beta 
\gamma } H_{\gamma }{}^{\kappa \lambda } H_{\theta }{}^{\mu 
\nu } H_{\kappa \mu }{}^{\tau } H_{\lambda \nu \tau } \nn\\&&+ 
\frac{1}{48} H_{\alpha \beta }{}^{\theta } H^{\alpha \beta 
\gamma } H_{\gamma }{}^{\kappa \lambda } H_{\theta }{}^{\mu 
\nu } H_{\kappa \lambda }{}^{\tau } H_{\mu \nu \tau } - 2 H_{\alpha }{}^{\theta \kappa } 
H^{\alpha \beta \gamma }R_{\beta \theta }{}^{\lambda 
\mu }R_{\gamma \lambda \kappa \mu } -  H_{\alpha \beta 
}{}^{\theta } H^{\alpha \beta \gamma }R_{\gamma 
}{}^{\kappa \lambda \mu }R_{\theta \lambda \kappa 
\mu } \nn\\&&+ 2 H^{\alpha 
\beta \gamma } H^{\theta \kappa \lambda }R_{\alpha 
\beta \theta }{}^{\mu }R_{\gamma \mu \kappa \lambda 
} - 2 H_{\alpha }{}^{\theta \kappa } H^{\alpha \beta \gamma } 
R_{\beta }{}^{\lambda }{}_{\gamma }{}^{\mu } 
R_{\theta \lambda \kappa \mu }+ \frac{1}{4} H^{\alpha \beta \gamma } H^{\theta \kappa 
\lambda } \nabla_{\gamma }H_{\kappa \lambda \mu } 
\nabla_{\theta }H_{\alpha \beta }{}^{\mu } \nn\\&& + \frac{1}{2} 
H_{\alpha }{}^{\theta \kappa } H^{\alpha \beta \gamma } 
\nabla_{\kappa }H_{\theta \lambda \mu } \nabla^{\mu }H_{\beta 
\gamma }{}^{\lambda } + H_{\alpha }{}^{\theta \kappa } 
H^{\alpha \beta \gamma } \nabla_{\mu }H_{\gamma \kappa 
\lambda } \nabla^{\mu }H_{\beta \theta }{}^{\lambda }\Big]\,.\labell{finalB}
\eeqa
Note that the Riemann-cubed terms and the six-$H$ terms with a coefficient of $1/12$ are invariant under field redefinitions, so these terms appear in three different forms of the couplings in Eq. (\ref{S2f}), (\ref{Sb}), and \reef{finalB}. We expect the above action to be the simplest form of the couplings in the bosonic string theory at order $\alpha'^2$, such that its corresponding couplings at order $\alpha'$ are described by the Meissner action  \reef{Mis}.

The heterotic string theory has 496 gauge vector fields as well as NS-NS fields. This theory exhibits a gauge anomaly, which can be canceled by assuming the gauge group to be $SO(32)$ or $E_8\times E_8$, and by allowing the $B$-field to undergo non-standard gauge transformations and local Lorentz transformations \cite{Green:1984sg}. In this paper, we specifically consider the case of zero gauge field. Under the non-standard local Lorentz transformation for the $B$-field, the $B$-field strength in Eq. (\ref{action1}), (\ref{S1bf}), or (\ref{Mis}) needs to be replaced by a new field strength that is invariant under these non-standard local Lorentz transformations, \ie
\beqa
H_{\mu\nu\alpha}&\rightarrow H_{\mu\nu\alpha}+\frac{3}{2}\alpha'\Omega_{\mu\nu\alpha}\,,\labell{replace}
\eeqa
where the  Chern-Simons three-form $\Omega$   is 
\beqa
\Omega_{\mu\nu\alpha}&=&\omega_{[\mu {\mu_1}}{}^{\nu_1}\prt_\nu\omega_{\alpha] {\nu_1}}{}^{\mu_1}+\frac{2}{3}\omega_{[\mu {\mu_1}}{}^{\nu_1}\omega_{\nu {\nu_1}}{}^{\alpha_1}\omega_{\alpha]{\alpha_1}}{}^{\mu_1}\,\,;\,\,\,\omega_{\mu {\mu_1}}{}^{\nu_1}=e^\nu{}_{\mu_1}\nabla_\mu e_\nu{}^{\nu_1} \,,
\eeqa
 where $e_\mu{}^{\mu_1}e_\nu{}^{\nu_1}\eta_{\mu_1\nu_1}=G_{\mu\nu}$. Our index convention is that $\mu, \nu, \dots$ are the indices of the curved spacetime, and $\mu_1, \nu_1, \dots$ are the indices of the flat tangent spaces. 
 
 The replacement of Eq. (\ref{replace}) into Eq. (\ref{action1}) yields the following terms at order $\alpha'$, which are parity-odd, and at order $\alpha'^2$, which are parity-even:
\beqa
\bS^{(1)}_O&=&-\frac{2\alpha'a_1}{\kappa^2}\int d^{10} x \sqrt{-G} \,e^{-2\Phi}\left(-2H_{\mu\nu\alpha}\Omega^{\mu\nu\alpha}\right)\nn\\
\bS^{(2)}_{1e}&=&-\frac{2\alpha'^2a_1^2}{\kappa^2}\int d^{10} x \sqrt{-G} \,e^{-2\Phi}\left(-12\Omega_{\mu\nu\alpha}\Omega^{\mu\nu\alpha}\right)
\labell{CS}
\eeqa
The replacement of Eq. (\ref{replace}) into Eq. (\ref{Mis}) produces the following terms at order $\alpha'^2$, which are parity-odd:
\beqa
\bS_{1O}^{(2)}&\!\!\!\!\!=\!\!\!\!\!&-\frac{2\alpha'^2 a_1^2}{\kappa^2}\int d^{10} x\sqrt{-G} e^{-2\Phi}\Big[-4 H^{\alpha \beta \gamma } R \Omega _{\alpha \beta 
\gamma } - 12 H^{\alpha \beta \gamma } R_{\beta \gamma 
\delta \epsilon } \Omega _{\alpha }{}^{\delta \epsilon } + 24 
H_{\alpha }{}^{\gamma \delta } R^{\alpha \beta } 
\Omega _{\beta \gamma \delta } \nn\\&&\qquad\qquad+ 2 H_{\alpha }{}^{\delta 
\epsilon } H^{\alpha \beta \gamma } H_{\beta \delta 
}{}^{\varepsilon } \Omega _{\gamma \epsilon \varepsilon } - 6 
H_{\alpha \beta }{}^{\delta } H^{\alpha \beta \gamma } 
H_{\gamma }{}^{\epsilon \varepsilon } \Omega _{\delta \epsilon 
\varepsilon } + \frac{1}{3} H_{\alpha \beta \gamma } H^{\alpha 
\beta \gamma } H^{\delta \epsilon \varepsilon } \Omega 
_{\delta \epsilon \varepsilon } \nn\\&&\qquad\qquad- 16 H^{\beta \gamma \delta } 
\Omega _{\beta \gamma \delta } \nabla_{\alpha 
}\nabla^{\alpha }\Phi + 16 H^{\beta \gamma \delta } \Omega _{
\beta \gamma \delta } \nabla_{\alpha }\Phi \nabla^{\alpha 
}\Phi + 48 H_{\alpha }{}^{\gamma \delta } \Omega _{\beta 
\gamma \delta } \nabla^{\beta }\nabla^{\alpha }\Phi  \Big],\labell{Misodd}
\eeqa
where $a_1=1/8$. 

There are no independent odd-parity couplings at order $\alpha'$ that do not involve $\Omega$. However, at order $\alpha'^2$, there are 13 independent odd-parity couplings in closed spacetime manifolds that do not involve $\Omega$ \cite{Garousi:2023kxw}.
It has been shown in \cite{Garousi:2023pah} that the parity-odd term in Eq. (\ref{CS}) is invariant under T-duality. However, it has been observed in \cite{Garousi:2023kxw} that the aforementioned parity-odd couplings at order $\alpha'^2$ in \reef{Misodd} are not invariant under T-duality. Therefore, T-duality necessitates the addition of the following parity-odd terms to the theory \cite{Garousi:2023kxw}:
\beqa
\bS_{2O}^{(2)}&=&-\frac{2\alpha'^2 a_1^2}{\kappa^2}\int d^{10} x\sqrt{-G} e^{-2\Phi}\Big[ 4H^{\alpha \beta \gamma } H^{\delta \epsilon 
\varepsilon } R_{\gamma \epsilon \varepsilon \mu } 
\nabla_{\beta }H_{\alpha \delta }{}^{\mu } -  2 
H_{\alpha }{}^{\delta \epsilon } H^{\alpha \beta \gamma } 
R_{\delta \varepsilon \epsilon \mu } \nabla^{\mu 
}H_{\beta \gamma }{}^{\varepsilon }\nn\\&&\qquad\qquad\qquad\qquad\qquad\quad  -  \frac{1}{2} H_{
\alpha }{}^{\delta \epsilon }H^{\alpha \beta \gamma } 
H_{\beta \delta }{}^{\varepsilon } H_{\gamma }{}^{\mu \zeta } 
\nabla_{\zeta }H_{\epsilon \varepsilon \mu }\Big]\,.\labell{finalodd}
\eeqa
Note that the dilaton only appears as an overall factor, so we do not attempt further field redefinitions to simplify it.

It has been observed in \cite{Garousi:2023kxw} that the parity-even couplings at order $\alpha'^2$ in Eq. (\ref{CS}) are also not invariant under T-duality. Therefore, T-duality necessitates the addition of other parity-even independent couplings that do not involve $\Omega$. There are 60 such couplings \cite{Garousi:2019cdn}. By adding these couplings with unfixed coefficients to the $\Omega^2$-term in Eq. (\ref{CS}), and utilizing the Meissner action in Eq. (\ref{Mis}), along with the odd-parity coupling in Eq. (\ref{CS}) for the couplings at order $\alpha'$, T-duality fixes the 60 couplings to be  the following \cite{Garousi:2023kxw}:
\beqa
\bS_{2e}^{(2)}&\!\!\!\!\!\!\!=\!\!\!\!\!\!\! &\frac{2\alpha'^2 a_1^2 }{\kappa^2}\int d^{10} x\sqrt{-G} e^{-2\Phi}\Big[\frac{1}{12} H_{\alpha }{}^{\delta \epsilon } H^{\alpha \beta 
\gamma } H_{\beta \delta }{}^{\zeta } H_{\gamma }{}^{\iota 
\kappa } H_{\epsilon \iota }{}^{\mu } H_{\zeta \kappa \mu }\labell{final}\\&& -  
\frac{1}{80} H_{\alpha \beta }{}^{\delta } H^{\alpha \beta 
\gamma } H_{\gamma }{}^{\epsilon \zeta } H_{\delta }{}^{\iota 
\kappa } H_{\epsilon \zeta }{}^{\mu } H_{\iota \kappa \mu } + 
\frac{1}{80} H_{\alpha \beta }{}^{\delta } H^{\alpha \beta 
\gamma } H_{\gamma }{}^{\epsilon \zeta } H_{\delta \epsilon 
}{}^{\iota } H_{\zeta }{}^{\kappa \mu } H_{\iota \kappa \mu }\nn\\&& 
-  \frac{2}{5} H_{\alpha }{}^{\delta \epsilon } H^{\alpha 
\beta \gamma } H_{\beta }{}^{\zeta \iota } H_{\delta \zeta 
}{}^{\kappa } R_{\gamma \epsilon \iota \kappa } + 2 H_{
\alpha }{}^{\delta \epsilon } H^{\alpha \beta \gamma } 
R_{\beta }{}^{\zeta }{}_{\delta }{}^{\iota } 
R_{\gamma \zeta \epsilon \iota }- 2 H^{\alpha \beta 
\gamma } H^{\delta \epsilon \zeta } R_{\alpha \beta 
\delta }{}^{\iota } R_{\gamma \iota \epsilon \zeta } \nn\\&&-  \frac{1}{40} 
H_{\alpha \beta }{}^{\delta } H^{\alpha \beta \gamma } 
H_{\epsilon \zeta }{}^{\kappa } H^{\epsilon \zeta \iota } 
R_{\gamma \iota \delta \kappa }  - 
2 H_{\alpha }{}^{\delta \epsilon } H^{\alpha \beta \gamma } 
R_{\beta }{}^{\zeta }{}_{\delta }{}^{\iota } 
R_{\gamma \iota \epsilon \zeta } + 2 H_{\alpha 
}{}^{\delta \epsilon } H^{\alpha \beta \gamma } 
R_{\beta }{}^{\zeta }{}_{\gamma }{}^{\iota } 
R_{\delta \zeta \epsilon \iota }\nn\\&& -  \frac{1}{10} 
H_{\alpha \beta }{}^{\delta } H^{\alpha \beta \gamma } 
H_{\gamma }{}^{\epsilon \zeta } H_{\epsilon }{}^{\iota \kappa } 
R_{\delta \iota \zeta \kappa } -  \frac{8}{5} 
H_{\alpha }{}^{\delta \epsilon } H^{\alpha \beta \gamma } 
H_{\beta \delta }{}^{\zeta } H_{\gamma }{}^{\iota \kappa } 
R_{\epsilon \iota \zeta \kappa }\nn\\&& -  \frac{1}{20} 
H_{\alpha \beta }{}^{\delta } H^{\alpha \beta \gamma } 
H_{\gamma }{}^{\epsilon \zeta } H_{\delta }{}^{\iota \kappa } 
R_{\epsilon \iota \zeta \kappa } -  \frac{1}{20} 
H_{\alpha }{}^{\gamma \delta } H_{\beta \gamma }{}^{\epsilon } 
H_{\delta }{}^{\zeta \iota } H_{\epsilon \zeta \iota } 
\nabla^{\beta }\nabla^{\alpha }\Phi\nn\\&& -  \frac{1}{10} H_{\gamma 
\delta }{}^{\zeta } H^{\gamma \delta \epsilon } 
R_{\alpha \epsilon \beta \zeta } \nabla^{\beta 
}\nabla^{\alpha }\Phi -  \frac{2}{5} H_{\alpha }{}^{\gamma 
\delta } H_{\gamma }{}^{\epsilon \zeta } R_{\beta 
\epsilon \delta \zeta } \nabla^{\beta }\nabla^{\alpha }\Phi\nn\\&& + 
\frac{1}{5} H_{\alpha }{}^{\gamma \delta } H_{\beta 
}{}^{\epsilon \zeta } R_{\gamma \epsilon \delta \zeta 
} \nabla^{\beta }\nabla^{\alpha }\Phi -  \frac{1}{10} 
\nabla^{\beta }\nabla^{\alpha }\Phi \nabla_{\epsilon 
}H_{\beta \gamma \delta } \nabla^{\epsilon }H_{\alpha 
}{}^{\gamma \delta } \nn\\&&-  \frac{1}{20} H_{\beta \gamma 
}{}^{\epsilon } H^{\beta \gamma \delta } H_{\delta }{}^{\zeta 
\iota } \nabla^{\alpha }\Phi \nabla_{\iota }H_{\alpha 
\epsilon \zeta } + \frac{1}{40} H_{\alpha }{}^{\beta \gamma } 
H_{\delta \epsilon }{}^{\iota } H^{\delta \epsilon \zeta } 
\nabla^{\alpha }\Phi \nabla_{\iota }H_{\beta \gamma \zeta } \nn\\&&- 
 \frac{1}{20} H_{\alpha }{}^{\delta \epsilon } H^{\alpha \beta 
\gamma } \nabla_{\iota }H_{\delta \epsilon \zeta } 
\nabla^{\iota }H_{\beta \gamma }{}^{\zeta } -  \frac{1}{5} H_{
\alpha }{}^{\delta \epsilon } H^{\alpha \beta \gamma } 
\nabla_{\zeta }H_{\gamma \epsilon \iota } \nabla^{\iota 
}H_{\beta \delta }{}^{\zeta }\nn\\&& + \frac{1}{5} H_{\alpha 
}{}^{\delta \epsilon } H^{\alpha \beta \gamma } \nabla_{\iota 
}H_{\gamma \epsilon \zeta } \nabla^{\iota }H_{\beta \delta 
}{}^{\zeta }\!\! +\!\! \frac{1}{5} H_{\alpha \beta }{}^{\delta } 
H^{\alpha \beta \gamma } \nabla_{\zeta }H_{\delta \epsilon 
\iota } \nabla^{\iota }H_{\gamma }{}^{\epsilon \zeta }\!\!-  
\!\!\frac{3}{40} H_{\alpha \beta }{}^{\delta } H^{\alpha \beta 
\gamma } \nabla_{\iota }H_{\delta \epsilon \zeta } 
\nabla^{\iota }H_{\gamma }{}^{\epsilon \zeta }
\Big].\nn
\eeqa
Note that the aforementioned couplings do not include the Riemann-cubed terms, which are consistent with the sphere-level S-matrix elements in string theory \cite{Metsaev:1986yb}.

Since the aforementioned couplings involve derivatives of the dilaton, we perform field redefinitions that only involve terms at order $\alpha'^2$ to simplify these couplings. We have succeeded in transforming the aforementioned 24 couplings into the following 8 couplings:
\beqa
\bS_{2e}^{(2)}&\!\!\!\!\!=-\!\!\!\!\!&\frac{2\alpha'^2 a_1^2}{\kappa^2}\int d^{10} x\sqrt{-G} e^{-2\Phi}\Big[ -\frac{1}{12} H_{\alpha }{}^{\theta \kappa } H^{\alpha \beta 
\gamma } H_{\beta \theta }{}^{\lambda } H_{\gamma }{}^{\mu \nu 
} H_{\kappa \mu }{}^{\tau } H_{\lambda \nu \tau }\nn\\&& +  
\frac{1}{4} H_{\alpha \beta }{}^{\theta } H^{\alpha \beta 
\gamma } H_{\gamma }{}^{\kappa \lambda } H_{\theta }{}^{\mu 
\nu } H_{\kappa \mu }{}^{\tau } H_{\lambda \nu \tau } - 2 
H_{\alpha }{}^{\theta \kappa } H^{\alpha \beta \gamma } 
R_{\beta \theta }{}^{\lambda \mu } R_{\gamma 
\lambda \kappa \mu } + 2 H^{\alpha \beta \gamma } H^{\theta 
\kappa \lambda } R_{\alpha \beta \theta }{}^{\mu } 
R_{\gamma \mu \kappa \lambda } \nn\\&&- 2 H_{\alpha 
}{}^{\theta \kappa } H^{\alpha \beta \gamma } R_{\beta 
}{}^{\lambda }{}_{\gamma }{}^{\mu } R_{\theta \lambda 
\kappa \mu } - \frac{1}{2} H^{\alpha \beta \gamma } H^{\theta 
\kappa \lambda } \nabla_{\alpha }H_{\theta \beta }{}^{\mu } 
\nabla_{\lambda }H_{\gamma \kappa \mu }+  \frac{1}{4} 
H_{\alpha }{}^{\theta \kappa } H^{\alpha \beta \gamma } 
\nabla_{\kappa }H_{\theta \lambda \mu } \nabla^{\mu }H_{\beta 
\gamma }{}^{\lambda }\nn\\&& +  H_{\alpha }{}^{\theta \kappa } 
H^{\alpha \beta \gamma } \nabla_{\mu }H_{\gamma \kappa 
\lambda } \nabla^{\mu }H_{\beta \theta }{}^{\lambda }\Big]\,.\labell{finaleven}
\eeqa
In the rest of this paper, our objective is to examine the couplings in Eq. (\ref{finalB}) within the bosonic string theory, as well as the couplings in Eq. (\ref{finalodd}) and Eq. (\ref{finaleven}) within the heterotic theory. We achieve this in Section 2 by studying the cosmological reduction of these couplings and demonstrating their invariance under $O(d,d)$ transformations. Furthermore, in Section 3, we compare these couplings with the 4-point sphere-level S-matrix elements and find exact consistency between the field theory and string theory S-matrix elements. We have used the "xAct" package \cite{Nutma:2013zea} for performing our calculations in this paper.

\section{Consistency with cosmological symmetry} 

When applying cosmological reduction to the classical effective action, the resulting one-dimensional effective action should exhibit $O(d,d,\MR)$ symmetry \cite{Sen:1991zi, Hohm:2014sxa}.
By utilizing various one-dimensional equations of motion, including the equation of motion for the lapse function, and employing integration by parts, it has been demonstrated in \cite{Hohm:2015doa, Hohm:2019jgu} that the cosmological reduction of the classical effective action of string theory at order $\alpha'$ and higher can be expressed in a scheme where only the first time-derivative of the generalized metric ${\cal S}$ appears.  The couplings involving $\mathrm{tr}(\dS^2)$ can be eliminated through a lapse function transformation, and the trace of an odd number of first derivatives of $\cS$ vanishes \cite{Hohm:2019jgu}. Consequently, the one-dimensional action can be expressed in a canonical form as the following expansion \cite{Hohm:2015doa, Hohm:2019jgu}:
\beqa
S_{\rm eff}^c&=&S_0^c-\frac{2}{\kappa^2}\int dt e^{-\phi}\bigg(\alpha' c_{2,0}\tr(\dS^4)+\alpha'^2c_{3,0}\tr(\dS^6)\nn\\
&&\qquad\quad\qquad\quad+\alpha'^3[c_{4,0}\tr(\dS^8)+c_{4,1}(\tr(\dS^4))^2]\nn\\
&&\qquad\quad\qquad\quad+\alpha'^4[c_{5,0}\tr(\dS^{10})+c_{5,1}\tr(\dS^6)\tr(\dS^4)]
+\cdots\bigg)\,.\labell{cosm}
\eeqa
In the above action, the coefficients $c_{m,n}$ depend on the specific theory. For example, $c_{2,0}$ and $c_{3,0}$ are non-zero for the bosonic string theory, while these numbers are zero for the superstring theory. The lapse function in the aforementioned action is set to $n=1$. By examining the cosmological reduction of only the pure gravity components of the couplings in various theories, one can determine the coefficients $c_{m,n}$. These coefficients up to order $\alpha'^3$ have been determined in \cite{Codina:2021cxh}.

Using the equations of motion and integration by parts is equivalent to employing the most general field redefinitions and disregarding total derivative terms. Therefore, any proposal for the classical effective action of string theory should be expressed in the canonical form given by Eq. (\ref{cosm}) after applying cosmological reduction, utilizing the most general one-dimensional field redefinitions, including the lapse function, and neglecting the one-dimensional total derivative terms. The NS-NS couplings at order $\alpha'^3$ in superstring theory, which were derived in \cite{Garousi:2020gio}, have been shown to exhibit a cosmological reduction that can be written in the aforementioned canonical form \cite{Garousi:2021ikb}. Furthermore, the couplings at order $\alpha'^2$ in the bosonic string theory, as presented in Eq. (\ref{S2f}), and their corresponding couplings at order $\alpha'$, which are described by the Metsaev-Tseytlin action in Eq. (\ref{S1bf}), have been shown to undergo a cosmological reduction that can be expressed in the canonical form \cite{Garousi:2021ocs}. We have observed that the cosmological couplings at order $\alpha'^2$ in bosonic string theory, as described in Eq. (\ref{finalB}), and their corresponding couplings at order $\alpha'$, which are governed by the Meissner action in Eq. (\ref{Mis}), exhibit the same canonical form. This result is not surprising, as the two action forms are connected solely through field redefinitions. We have also performed calculations for the heterotic couplings presented in Eq. (\ref{CS}), Eq. (\ref{Misodd}), Eq. (\ref{finalodd}), and Eq. (\ref{finaleven}), and obtained similar outcomes.

If the classical effective action has the following $\alpha'$-expansion:
\beqa
\bS_{\rm eff} = \sum^\infty_{m=0}\alpha'^m\bS_m = \bS_0+\alpha' \bS_1 +\alpha'^2 \bS_2+\alpha'^3 \bS_3+\cdots\,. \labell{seff}
\eeqa
The corresponding cosmological reduction would be:
\beqa
\bS^c_{\rm eff}(\psi) = \sum^\infty_{m=0}\alpha'^m\bS^c_m(\psi) = \bS^c_0(\psi)+\alpha' \bS^c_1(\psi) +\alpha'^2 \bS^c_2(\psi)+\alpha'^3 \bS^c_3(\psi)+\cdots \,.\labell{seffc}
\eeqa
Here, $\psi$ collectively represents the one-dimensional functions. The cosmological reduced actions $\bS^c_m(\psi)$ are not invariant under $O(d,d)$-transformations unless one uses appropriate one-dimensional field redefinitions $\psi\rightarrow \psi+\alpha'\delta\psi^{(1)}+\alpha'^2\delta\psi^{(2)}+\cdots$, and includes total derivative terms $\cJ$. The $O(d,d)$-invariant action $S^c_{\rm eff}$ should also have the following $\alpha'$-expansion:
\beqa
S^c_{\rm eff}(\psi)&=&\sum^\infty_{m=0}\alpha'^m S^c_m(\psi)=S^c_0(\psi)+\alpha' S^c_1(\psi) +\alpha'^2 S^c_2(\psi)+\alpha'^3 S^c_3(\psi)+\cdots \,.\labell{seffc1}
\eeqa
If one uses the replacement $\psi\rightarrow \psi+\alpha'\delta\psi^{(1)}+\alpha'^2\delta\psi^{(2)}+\cdots$ in the actions $\bS^c_m$ and then expands the actions in powers of $\alpha'$, one would find the following relations between the expansions and the $O(d,d)$-invariant actions up to order $\alpha'^2$:
\beqa
S_0^c&=&\!\!\bS_0^c(\psi)+\cJ_0,\nn\\
S_1^c&=&\!\!\bS_1^c(\psi)+\delta_1\!\!\bS_0^c(\psi)+\cJ_1,\labell{S012}\\
S_2^c&=&\!\!\bS_2^c(\psi)+\delta_1\!\!\bS_1^c(\psi)+\frac{1}{2}\delta_1^2\!\!\bS_0^c(\psi)+\delta_2\!\!\bS_0^c(\psi)+\cJ_2,\nn
\eeqa
where $\delta_i$ on the action indicates that the action contains the first-order perturbation $\delta\psi^{(i)}$, and $\delta_1^2$ on the action means the action contains the second-order perturbation $\delta\psi^{(1)}\delta\psi^{(1)}$. We have also used the $\alpha'$-expansion of the total derivative terms as $\cJ = \sum_{m=0}^\infty \alpha'^m \cJ_m$.

Since one-dimensional field redefinitions involve the lapse function, and the lapse function appears nontrivially in the cosmological reduced actions $\bS^c_m$, it is difficult to use the above relations to determine the $O(d,d)$-invariant actions $S^c_m$. However, $O(d,d)$-covariance requires this function to appear in the $O(d,d)$-invariant form of the action by replacing the measure of the integral as $dt\rightarrow dt/n^{(2k-1)}$, where $2k$ is the number of time-derivatives in the action. For example, the lapse function in $S_1^c$ is modified as $dt\rightarrow dt/n^3$. Hence, it is appropriate to relate the perturbation of the cosmological reduced action to the perturbation of the $O(d,d)$-invariant actions. This allows us to find the relations between the expansions of $O(d,d)$-invariant action $S^c_m$ in terms of perturbations of $O(d,d)$-invariant actions $S^c_p$ for $p<m$. These relations, up to order $\alpha'^2$, are as follows  \cite{Garousi:2021ocs}:
 \beqa
S_0^c&=&\!\!\bS_0^c(\psi)+\cJ_0,\nn\\
S_1^c&=&\!\!\bS_1^c(\psi)+\delta_1S_0^c(\psi)+\cJ_1,\labell{S0122}\\
S_2^c&=&\!\!\bS_2^c(\psi)+\delta_1S_1^c(\psi)-\frac{1}{2}\delta_1^2S_0^c(\psi)+\delta_2S_0^c(\psi)+\cJ_2\,.\nn
\eeqa
 In this section, we are going to demonstrate that when the cosmological reduction of the effective  actions up to order $\alpha'^2$ is inserted into the above expressions, they satisfy the $O(d,d)$ symmetry and can be written in the minimal form of  \reef{cosm}.

\subsection{Cosmological symmetry at the leading order}

In this subsection, we review the $O(d,d)$ or cosmological symmetry of the leading-order action \cite{Veneziano:1991ek,Meissner:1991zj,Maharana:1992my,Meissner:1996sa}. When fields depend only on time, it is possible to write the metric, $B$-field, and dilaton using the gauge symmetries as follows:
 \beqa
G_{\mu\nu}=\left(\matrix{-n^2(t)& 0&\cr 0&G_{ij}(t)&}\right),\, B_{\mu\nu}= \left(\matrix{0&0\cr0&B_{ij}(t)&}\right),\,  2\Phi=\phi+\frac{1}{2}\log\det(G_{ij})\,,\labell{creduce}\eeqa
where the lapse function $n(t)$ can also be fixed to $n=1$. The cosmological reduction of the action in  \reef{action1} then becomes:
\beqa
\bS_0^c&=&-\frac{2}{\kappa^2}\int dt e^{-\phi}\Bigg[\frac{1}{4}\dB_{ij}\dB^{ij}-\frac{3}{4}\dG_{ij}\dG^{ij}-G^{ij}\dG_{ij}\dP-\dP^2+G^{ij}\ddot{G}_{ij}\Bigg]\,,
\eeqa
where $\dG^{ij}\equiv G^{ik}G^{il}\dG_{kl}$. Removing a  total derivative term,
one can write $\bS_0^c$ as
\beqa
S_0^c&=&\bS_0^c+\cJ_0\,=\,-\frac{2}{\kappa^2}\int dt e^{-\phi}\Bigg[\frac{1}{4}\dB_{ij}\dB^{ij}+\frac{1}{4}\dG_{ij}\dG^{ij}-\dP^2\Bigg]\,.\labell{S0c}
\eeqa
Using the generalized metric $\cS$ which is defined as 
\beqa
\cS\equiv \eta \left(\matrix{G^{-1}& -G^{-1}B&\cr BG^{-1}&G-BG^{-1}B&}\!\!\!\!\!\right)\labell{S}\,,%=\left(\matrix{BG^{-1}& G-BG^{-1}B&\cr G^{-1}&-G^{-1}B&}\right)
\eeqa
where $\eta$ is the  metric of the $O(d,d)$ group which in the non-diagonal form is 
\beqa
\eta&=& \left(\matrix{0& 1&\cr 1&0&}\!\!\!\!\!\right),
\eeqa
one can write the  action \reef{S0c} as
\beqa
S_0^c&=&-\frac{2}{\kappa^2}\int dt e^{-\phi}\Bigg[-\dP^2-\frac{1}{8}\tr(\dS^2)\Bigg]\,.\labell{S0}
\eeqa
This  action is invariant under global $O(d,d)$ transformations. The lapse function can be incorporated into the action \reef{S0} by replacing $dt$ with $dt/n$.
 
\subsection{Cosmological symmetry at order $\alpha'$}

In this subsection, we will demonstrate that the cosmological reduction of the couplings in \reef{Mis} and the odd-parity couplings at order $\alpha'$ in \reef{CS} can be expressed in an $O(d,d)$-invariant form by employing suitable one-dimensional field redefinitions and incorporating total derivative terms.

Using the reductions in \reef{creduce}, we obtain the following cosmological reduction for the actions at order $\alpha'$:
\beqa
 \bS_1^c&\!\!\!\!\!=\!\!\!\!\!&-\frac{2a_1}{\kappa^2}\int dt e^{-\phi}\Bigg[- \frac{3}{8} \dB_{i}{}^{k} \dB^{ij} \dB_{j}{}^{l} \dB_{kl} -  
\frac{1}{16} \dB_{ij} \dB^{ij} \dB_{kl} \dB^{kl} + \frac{1}{4} 
\dB^{ij} \dB^{kl} \dG_{ik} \dG_{jl} + \frac{1}{2} \dB_{i}{}^{k} 
\dB^{ij} \dG_{j}{}^{l} \dG_{kl} \nn\\&&-  \frac{7}{8} \dG_{i}{}^{k} 
\dG^{ij} \dG_{j}{}^{l} \dG_{kl} -  \frac{1}{8} \dB_{ij} \dB^{ij} 
\dG_{kl} \dG^{kl} + \frac{7}{16} \dG_{ij} \dG^{ij} \dG_{kl} 
\dG^{kl} + \dB_{i}{}^{k} \dB^{ij} \dG_{jk} \dP + \frac{1}{2} 
\dB_{ij} \dB^{ij} \dP^2 \nn\\&&-  \frac{1}{2} \dG_{ij} \dG^{ij} 
\dP^2 -  \dP^4 + \dG_{i}{}^{k} \dG^{ij} \ddot{G}_{jk} + 2 
\dP^2 \ddot{\phi} + \dG_{i}{}^{k} \dG^{ij} \dG_{jk} \dG^i{}_i -  
\frac{1}{2} \dG_{ij} \dG^{ij} \dP \dG^i{}_i -  \dP^3 \dG^i{}_i \nn\\&&-  
\dG^{ij} \ddot{G}_{ij} \dG^i{}_i + 2 \dP \ddot{\phi} \dG^i{}_i -  \frac{3}{4} 
\dG_{ij} \dG^{ij} (\dG^i{}_i)^2 -  \frac{1}{2} \dP^2 (\dG^i{}_i)^2 + 
\frac{1}{2} \ddot{\phi} (\dG^i{}_i)^2-  \frac{1}{4} \dP( \dG^i{}_i)^3\nn\\&&  -  
\frac{1}{2} \dG_{ij} \dG^{ij} \ddot{G}^j{}_j + \dP^2 \ddot{G}^j{}_j + \dP \dG^i{}_i \ddot{G}^j{}_j 
+ \frac{3}{4} (\dG^i{}_i)^2 \ddot{G}^j{}_j- \dB^{ij} \dG_{i}{}^{k} \ddot{G}_{jk}\Bigg]\,,\labell{cS3}
 \eeqa
where we have used the gauge $n=1$. It is important to note that the last term above corresponds to the cosmological reduction of the odd-parity term in  \reef{CS}. The above action is not invariant under $O(d,d)$ transformations. For example, it contains $\dG^i{}_i$, which is not invariant.  Some of the non-invariant terms are total derivative terms that should be eliminated. Additionally, the action expressed in terms of the variables $G_{ij}$, $B_{ij}$, and $\phi$ is not invariant. It should be invariant in terms of alternative variables that involve higher derivatives of $G_{ij}$, $B_{ij}$, and $\phi$. Therefore, one needs to employ one-dimensional higher-derivative field redefinitions.
 
To eliminate the total derivative terms from  \reef{cS3}, we incorporate additional terms with arbitrary coefficients, which are total derivatives at order $\alpha'$. The following total derivative terms are included:
\beqa
\cJ_1&=&-\frac{2}{\kappa^2}\int dt\frac{d}{dt}(e^{-\phi}\cI_1)\,,
\eeqa
where $\cI_1$ encompasses all possible terms at the three-derivative level, constructed from $\dP$, $\dB$, $\dG$, $\ddot{\phi}$, $\ddot{B}$, $\ddot{G}$, and so on.

One can perform a change of field variables in \reef{creduce} as follows:
\begin{eqnarray}
G_{ij} &\rightarrow &G_{ij}+\alpha' \delta G^{(1)}_{ij},\nn\\
B_{ij} &\rightarrow &B_{ij} + \alpha'\delta B^{(1)}_{ij},\nn\\
\phi &\rightarrow &\phi + \alpha'\delta\phi^{(1)},\nn\\
n &\rightarrow &n + \alpha' \delta n^{(1)}.\labell{gbpn}
\end{eqnarray}
Here, the matrices $\delta G^{(1)}_{ij}$, $\delta B^{(1)}_{ij}$, $\delta\phi^{(1)}$, and $\delta n^{(1)}$ consist of all possible terms at the 2-derivative level, constructed from $\dP$, $\dB$, $\dG$, $\ddot{\phi}$, $\ddot{B}$, $\ddot{G}$. When the field variables in $S_0^c$ are transformed according to the aforementioned field redefinitions, they give rise to certain couplings at order $\alpha'$ and higher. In this section, our focus lies on the resulting couplings at order $\alpha'$, \ie 
\beqa
\delta_1 S_0^c&=&-\frac{2}{\kappa^2}\int dt e^{-\Phi}\Bigg[\delta n^{(1)}\left(-\frac{1}{4}\dB_{ij}\dB^{ij}-\frac{1}{4}\dG_{ij}\dG^{ij}+\dP^2\right)\labell{dS0c}\\
&&+\delta \phi^{(1)}\left(-\frac{1}{4}\dB_{ij}\dB^{ij}-\frac{1}{4}\dG_{ij}\dG^{ij}+\dP^2\right)-2\dP\frac{d}{dt}\delta \phi^{(1)}\nn\\
&&+\delta G^{(1)}_{ij}\left(-\frac{1}{2}\dB_k{}^j\dB^{ki}-\frac{1}{2}\dG_k{}^j\dG^{ki}\right)+\frac{1}{2}\dG^{ij}\frac{d}{dt}\delta G^{(1)}_{ij}+\frac{1}{2}\dB^{ij}\frac{d}{dt}\delta B^{(1)}_{ij}\Bigg]\,,\nn
%&\equiv& -\frac{2\alpha' b_1}{\kappa^2}\int dt e^{-\Phi}{\cal K}_1\nn
\eeqa
where we have utilized the fact that the lapse function appears in the action \reef{S0c} through the replacement $dt\rightarrow dt/n$.

By employing the field redefinitions and incorporating the total derivative terms, one can express $\bS_1^c+\delta_1 S_0^c$ in various $O(d,d)$-invariant forms. However, if one wishes to write it in the canonical form given by  \reef{cosm}, certain constraints need to be imposed. These constraints require that $\bS_1^c+\delta_1 S_0^c$ only includes first derivative terms, does not contain derivatives of the dilaton, and does not include the trace of two $\dB$ or $\dG$ terms.
By imposing these constraints, one finds that the following field redefinitions can be employed:
 \beqa
\delta n^{(1)}&=&a_1\Big(- \frac{1}{4} \dB_{ij} \dB^{ij} - \frac{1}{4} \dG_{ij} \dG^{ij}+ \frac{1}{2}\dP^2\Big),\nn\\
\delta \phi^{(1)}&=&\frac{a_1}{2} \dP^2,\nn\\
\delta G^{(1)}_{ij}&=&-a_1( \dB_{i}{}^{k} -\dG_{i}{}^{k})(\dB_{jk} - \dG_{jk}),\nn\\
\delta B^{(1)}_{ij}&=&a_1\Big(\dB_{j}{}^{k} \dG_{ik} -  \dB_{i}{}^{k} \dG_{jk} \Big) .\labell{dG1dB1}
\eeqa
Then, up to some total derivative terms, the cosmological action at order $\alpha'$ can be expressed as follows: 
 \beqa
 S_1^c&=&\!\!\bS_1^c+\delta_1 S_0^c+\cJ_1\,=\,-\frac{2a_1}{\kappa^2}\int dt e^{-\phi}\Bigg[\frac{1}{8} \dB_{i}{}^{k} \dB^{ij} \dB_{j}{}^{l} \dB_{kl} -  
\frac{1}{4} \dB^{ij} \dB^{kl} \dG_{ik} \dG_{jl}\nn\\&&\qquad\qquad\qquad\quad + \frac{1}{2} 
\dB_{i}{}^{k} \dB^{ij} \dG_{j}{}^{l} \dG_{kl} + \frac{1}{8} 
\dG_{i}{}^{k} \dG^{ij} \dG_{j}{}^{l} \dG_{kl}\Bigg]\,.\labell{action2}
 \eeqa
The total derivative terms can be ignored as they have no impact on the calculations at order $\alpha'^2$.

Now, by utilizing the definition of the generalized metric in  \reef{S}, we can find the following expression:
\beqa
\tr(\dS^4)&=&2 \dB_{i}{}^{k} \dB^{ij} \dB_{j}{}^{l} \dB_{kl} - 4 \dB^{ij} 
\dB^{kl} \dG_{ik} \dG_{jl} + 8 \dB_{i}{}^{k} \dB^{ij} \dG_{j}{}^{l} 
\dG_{kl} + 2 \dG_{i}{}^{k} \dG^{ij} \dG_{j}{}^{l} \dG_{kl}\,.
\eeqa
 Using the aforementioned $O(d,d)$-invariant expressions, we can express \reef{action2} as follows: 
\beqa
 S_1^c&=&-\frac{2a_1}{16\kappa^2}\int dt e^{-\phi}\tr(\dS^4)\,.\labell{action3}
 \eeqa
This expression is consistent with the cosmological action in \reef{cosm} with $c_{2,0}=a_1/16=1/2^7$. For bosonic string theory, the calculations remain the same, except that there are no odd-parity terms in \reef{cS3} and in the $G_{ij}$-field redefinition in \reef{dG1dB1}. Therefore, for the bosonic theory, $c_{2,0}=a_1/16=1/2^6$. These results coincide with those found in \cite{Codina:2021cxh} through the study of pure gravity parts of the couplings. The lapse function can be incorporated into the above action by replacing $dt$ with $dt/n^3$.

\subsection{Cosmological symmetry at order $\alpha'^2$}

In this section, we are going to show that, up to one-dimensional field redefinitions and total derivative terms, the cosmological reduction of the couplings in Eq. (\ref{CS}), Eq. (\ref{Misodd}), Eq. (\ref{finalodd}), and Eq. (\ref{finaleven}) at order $\alpha'^2$ in the heterotic theory, and the cosmological reduction of the couplings in Eq. (\ref{finalB}) in the bosonic theory, can be written in the canonical form given by Eq. (\ref{cosm}).

By using the reductions in Eq. (\ref{creduce}), and the following reduction for $e_\mu{}^{\mu_1}$
\beqa
e_{\mu}{}^{\mu_1}=\left(\matrix{n(t)& 0&\cr 0&e_{i}{}^{i_1}(t)&}\right),\, \labell{creducee}
\eeqa
where $e_i{}^{i_1}e_j{}^{j_1}\eta_{i_1j_1}=G_{ij}$, we obtain the following cosmological reduction for the actions at order $\alpha'^2$ in the heterotic theory:
\beqa
\bS_{1e}^{c(2)}&=&\frac{2\alpha'^2a_1^2}{\kappa^2}\int dt e^{-\phi}\Bigg[\frac{1}{2} \dG^{ij} \dG^{kl} \ddot{G}_{ik} \ddot{G}_{jl} -  \frac{1}{2} 
\dG_{i}{}^{k} \dG^{ij} \ddot{G}_{j}{}^{l} \ddot{G}_{kl}\Bigg],\labell{eeOO}\\
\bS_{2e}^{c(2)}&=&\frac{2\alpha'^2a_1^2}{\kappa^2}\int dt e^{-\phi}\Bigg[- \frac{3}{4} \dB_{i}{}^{k} \dB^{ij} \dB_{j}{}^{l} \dB_{k}{}^{m} 
\dB_{l}{}^{n} \dB_{mn} -  \frac{1}{4} \dB_{ij} \dB^{ij} 
\dB_{k}{}^{m} \dB^{kl} \dB_{l}{}^{n} \dB_{mn}\nn\\&& -  \frac{3}{2} 
\dB_{i}{}^{k} \dB^{ij} \dB_{l}{}^{n} \dB^{lm} \dG_{jm} \dG_{kn} -  
\frac{3}{2} \dB_{i}{}^{k} \dB^{ij} \dB_{j}{}^{l} \dB^{mn} \dG_{km} 
\dG_{ln} + \frac{3}{4} \dB^{ij} \dB^{kl} \dG_{i}{}^{m} 
\dG_{j}{}^{n} \dG_{km} \dG_{ln}\nn\\&& + \frac{1}{2} \dB_{i}{}^{k} 
\dB^{ij} \dB_{j}{}^{l} \dB_{k}{}^{m} \dG_{l}{}^{n} \dG_{mn} -  
\dB^{ij} \dB^{kl} \dG_{ik} \dG_{j}{}^{m} \dG_{l}{}^{n} \dG_{mn} + 
\frac{1}{2} \dB_{i}{}^{k} \dB^{ij} \dG_{j}{}^{l} \dG_{k}{}^{m} 
\dG_{l}{}^{n} \dG_{mn} \nn\\&&-  \frac{1}{4} \dB_{i}{}^{k} \dB^{ij} 
\dB_{j}{}^{l} \dB_{kl} \dG_{mn} \dG^{mn} -  \frac{3}{2} \dB^{ij} 
\dB^{kl} \ddot{B}_{ik} \ddot{B}_{jl} -  \frac{3}{2} \dB_{i}{}^{k} \dB^{ij} 
\ddot{B}_{j}{}^{l} \ddot{B}_{kl} + 5 \dB_{i}{}^{k} \dB^{ij} \dB^{lm} \dG_{jl} 
\ddot{B}_{km}\nn\\&& -  \dB_{i}{}^{k} \dB^{ij} \dB_{j}{}^{l} \dG_{k}{}^{m} 
\ddot{B}_{lm} - 2 \dB^{ij} \dB^{kl} \dG_{i}{}^{m} \dG_{km} \ddot{G}_{jl} + 2 
\dB^{ij} \dB^{kl} \ddot{G}_{ik} \ddot{G}_{jl} + 2 \dB_{i}{}^{k} \dB^{ij} 
\ddot{G}_{j}{}^{l} \ddot{G}_{kl}\nn\\&& - 2 \dB_{i}{}^{k} \dB^{ij} \dG_{j}{}^{l} 
\dG_{l}{}^{m} \ddot{G}_{km} + 2 \dB^{ij} \dB^{kl} \dG_{ik} \dG_{j}{}^{m} 
\ddot{G}_{lm}\Bigg],\nn\\
\bS_{1O}^{c(2)}&=&\frac{2\alpha'^2a_1^2}{\kappa^2}\int dt e^{-\phi}\Bigg[- \frac{1}{2} \dB^{ij} \dG_{i}{}^{k} \dG_{lm} \dG^{lm} \ddot{G}_{jk} 
+ 2 \dB^{ij} \dG_{i}{}^{k} \dP^2 \ddot{G}_{jk} + 
2 \dB^{ij} \dG_{i}{}^{k} \dG_{k}{}^{l} \dP \ddot{G}_{jl} \nn\\&&
+  \dB^{ij} \dG_{i}{}^{k} \dG_{k}{}^{l} \dG_{l}{}^{m} 
\ddot{G}_{jm} - 3 \dB_{i}{}^{k} \dB^{ij} \dB_{j}{}^{l} 
\dG_{k}{}^{m} \ddot{G}_{lm} -  \frac{1}{2} \dB_{ij} \dB^{ij} 
\dB^{kl} \dG_{k}{}^{m} \ddot{G}_{lm}\Bigg],\nn\\
\bS_{2O}^{c(2)}&=&-\frac{2\alpha'^2a_1^2}{\kappa^2}\int dt e^{-\phi}\Bigg[-2 \dB_{i}{}^{k} \dB^{ij} \dB^{lm} \dG_{jl} \dG_{k}{}^{n} \dG_{mn} + 
\dB_{i}{}^{k} \dB^{ij} \dG_{j}{}^{l} \dG_{l}{}^{m} \ddot{B}_{km} \nn\\&& -  
\dB^{ij} \dB^{kl} \dG_{ik} \dG_{j}{}^{m} \ddot{B}_{lm}- 2 \dB_{i}{}^{k} 
\dB^{ij} \ddot{B}_{j}{}^{l} \ddot{G}_{kl} + 2 \dB_{i}{}^{k} \dB^{ij} \dB^{lm} 
\dG_{jl} \ddot{G}_{km} - 2 \dB_{i}{}^{k} \dB^{ij} \dB_{j}{}^{l} 
\dG_{k}{}^{m} \ddot{G}_{lm}\Bigg].\nn
\eeqa
 In the above action, we have chosen the gauge $n=1$.  It is obvious that the sum of the above actions, i.e., $ \!\!\bS_2^c=(\bS_{1e}^{c(2)}+\bS_{2e}^{c(2)}+\bS_{1O}^{c(2)}+\bS_{2O}^{c(2)})/\alpha'^2$, is not invariant under $O(d,d)$ transformations. It should be invariant after adding one-dimensional total derivative terms and using field redefinitions.

We add the following total derivative term at order $\alpha'^2$ to $\!\!\bS_2^c$:
\beqa
\cJ_2&=&-\frac{2}{\kappa^2}\int dt\frac{d}{dt}(e^{-\phi}\cI_2)\,,
\eeqa
where $\cI_2$ includes all possible terms at the five-derivative level, which are constructed from $\dP$, $\dB$, $\dG$, $\ddot{\phi}$, $\ddot{B}$, $\ddot{G}$, $\cdots$. 

We also make the following field redefinitions:
\begin{eqnarray}
G_{ij}&\rightarrow &G_{ij}+\alpha' \delta G^{(1)}_{ij}+\alpha'^2 \delta G^{(2)}_{ij},\nn\\
B_{ij}&\rightarrow &B_{ij}+ \alpha'\delta B^{(1)}_{ij}+ \alpha'^2\delta B^{(2)}_{ij},\nn\\
\phi &\rightarrow &\phi+ \alpha'\delta\phi^{(1)}+ \alpha'^2\delta\phi^{(2)},\nn\\
n &\rightarrow &n+ \alpha' \delta n^{(1)}+ \alpha'^2 \delta n^{(2)},\labell{gbpn2}
\end{eqnarray}
where the first-order perturbations $\delta G^{(1)}_{ij}$, $\delta B^{(1)}_{ij}$, $\delta \phi^{(1)}$, $\delta n^{(1)}$ are given in Eq. (\ref{dG1dB1}), and the second-order perturbations $\delta G^{(2)}_{ij}$, $\delta B^{(2)}_{ij}$, $\delta \phi^{(2)}$, $\delta n^{(2)}$ consist of all possible terms at the four-derivative level constructed from $\dP$, $\dB$, $\dG$, $\ddot{\phi}$, $\ddot{B}$, $\ddot{G}$, $\cdots$.

When the field variables in $S_0^c$ are changed according to the above field redefinitions, they produce two sets of couplings at order $\alpha'^2$. One set is produced by the first-order perturbations at order $\alpha'^2$, i.e.,
\beqa
\delta_2 S_0^c&=&-\frac{2}{\kappa^2}\int dt e^{-\phi}\Bigg[\delta n^{(2)}\left(-\frac{1}{4}\dB_{ij}\dB^{ij}-\frac{1}{4}\dG_{ij}\dG^{ij}+\dP^2\right)\labell{dS0c1}\\
&&+\delta \phi^{(2)}\left(-\frac{1}{4}\dB_{ij}\dB^{ij}-\frac{1}{4}\dG_{ij}\dG^{ij}+\dP^2\right)-2\dP\frac{d}{dt}\delta \phi^{(2)}\nn\\
&&+\delta G^{(2)}_{ij}\left(-\frac{1}{2}\dB_k{}^j\dB^{ki}-\frac{1}{2}\dG_k{}^j\dG^{ki}\right)+\frac{1}{2}\dG^{ij}\frac{d}{dt}\delta G^{(2)}_{ij}+\frac{1}{2}\dB^{ij}\frac{d}{dt}\delta B^{(2)}_{ij}\Bigg]\,,\nn
%&\equiv& -\frac{2\alpha' b_1}{\kappa^2}\int dt e^{-\Phi}{\cal K}_1\nn
\eeqa
which are similar to the first-order perturbation given in Eq. (\ref{dS0c}), and the other set is reproduced by the  second-order perturbations, i.e.,
\beqa
\frac{1}{2}\delta_1^2 S_0^c&=&-\frac{2}{\kappa^2}\int dt e^{-\Phi}\Bigg[\frac{1}{4} \frac{d}{dt}\delta B^{(1)}_{ij} \frac{d}{dt}\delta B^{(1)}{}^{ij} + \frac{1}{4} 
\frac{d}{dt}\delta G^{(1)}_{ij} \frac{d}{dt}\delta G^{(1)}{}^{ij} -  \frac{d}{dt}\delta \phi^{(1)} \frac{d}{dt}\delta \phi^{(1)}\nn\\&& -  \dB^{ij} \frac{d}{dt}\delta B^{(1)}_{i}{}^{k} \delta G^{(1)}_{jk} -  \dG^{ij} \frac{d}{dt}\delta G^{(1)}_{i}{}^{k} \delta G^{(1)}_{jk} + \frac{1}{4} \dB^{ij} \dB^{kl} 
\delta G^{(1)}_{ik} \delta G^{(1)}_{jl} + \frac{1}{4} \dG^{ij} \dG^{kl} 
\delta G^{(1)}_{ik} \delta G^{(1)}_{jl}\nn\\&& + \frac{1}{2} \dB_{i}{}^{k} 
\dB^{ij} \delta G^{(1)}_{j}{}^{l} \delta G^{(1)}_{kl} + \frac{1}{2} 
\dG_{i}{}^{k} \dG^{ij} \delta G^{(1)}_{j}{}^{l} \delta G^{(1)}_{kl} -  
\frac{1}{2} \dB^{ij} \frac{d}{dt}\delta B^{(1)}_{ij} \delta n^{(1)} -  \frac{1}{2} \dG^{ij} \frac{d}{dt}\delta G^{(1)}_{ij} \delta n^{(1)}\nn\\&& + 2 \dP \frac{d}{dt}\delta \Phi^{(1)} 
\delta n^{(1)} + \frac{1}{2} \dB_{i}{}^{k} \dB^{ij} \delta G^{(1)}_{jk} 
\delta n^{(1)} + \frac{1}{2} \dG_{i}{}^{k} \dG^{ij} \delta G^{(1)}_{jk} 
\delta n^{(1)} \nn\\&& +\Big( \frac{1}{4} \dB_{ij} \dB^{ij} + 
\frac{1}{4} \dG_{ij} \dG^{ij}  -  \dP^2\Big) 
\delta n^{(1)}\delta n^{(1)} + \Big(\frac{1}{8} \dB_{ij} \dB^{ij}  + \frac{1}{8} \dG_{ij} \dG^{ij} -  
\frac{1}{2} \dP^2\Big) \delta\phi^{(1)} \delta \phi^{(1)} \nn\\&& -  \frac{1}{2} \dG^{ij} \frac{d}{dt}\delta G^{(1)}_{ij} \delta \hi^{(1)} + 2 \dP \frac{d}{dt}\delta \phi^{(1)} \delta 
\phi^{(1)} + \frac{1}{2} \dB_{i}{}^{k} \dB^{ij} \delta G^{(1)}_{jk} 
\delta \phi^{(1)} + \frac{1}{2} \dG_{i}{}^{k} \dG^{ij} \delta  G^{(1)}_{jk} \delta \phi^{(1)}\nn\\&& -  \frac{1}{2} 
\dB^{ij} \frac{d}{dt}\delta B^{(1)}_{ij} \delta \phi^{(1)}+\Big( \frac{1}{4} \dB_{ij} \dB^{ij} + \frac{1}{4} \dG_{ij} \dG^{ij} -  \dP^2\Big) \delta n^{(1)} \delta \phi^{(1)} \Bigg]\,,\labell{d1}
\eeqa
where the first-order perturbations are given in Eq. (\ref{dG1dB1}).

When the field variables in the $O(d,d)$-invariant action at order $\alpha'$, as given by Eq. (\ref{action2}), are changed according to the field redefinition in Eq. (\ref{gbpn2}), one also finds the following couplings at order $\alpha'^2$ resulting from the first-order perturbation at order $\alpha'$:
\beqa
\delta_1 S_1^c&=&-\frac{2a_1}{\kappa^2}\int dt e^{-\Phi}\Bigg[- \dB^{ij} \dG_{i}{}^{k} \dG_{k}{}^{l} \frac{d}{dt}\delta B^{(1)}_{jl} + 
\frac{1}{2} \dB_{i}{}^{k} \dB^{ij} \dB_{j}{}^{l} \frac{d}{dt}\delta B^{(1)}_{kl} - 
 \frac{1}{2} \dB^{ij} \dG_{i}{}^{k} \dG_{j}{}^{l} \frac{d}{dt}\delta B^{(1)}_{kl} \nn\\&&
-  \frac{1}{2} \dB^{ij} \dB^{kl} \dG_{ik} \frac{d}{dt}\delta G^{(1)}_{jl} + 
\dB_{i}{}^{k} \dB^{ij} \dG_{j}{}^{l} \frac{d}{dt}\delta G^{(1)}_{kl} + \frac{1}{2} 
\dG_{i}{}^{k} \dG^{ij} \dG_{j}{}^{l} \frac{d}{dt}\delta G^{(1)}_{kl}\nn\\&& -  
\frac{1}{2} \dB^{ij} \dB^{kl} \dG_{i}{}^{m} \dG_{km} \delta 
G^{(1)}_{jl} -  \dB_{i}{}^{k} \dB^{ij} \dG_{j}{}^{l} \dG_{l}{}^{m} \delta 
G^{(1)}_{km} -  \frac{1}{2} \dB_{i}{}^{k} \dB^{ij} \dB_{j}{}^{l} 
\dB_{k}{}^{m} \delta G^{(1)}_{lm} \nn\\&&+ \dB^{ij} \dB^{kl} \dG_{ik} 
\dG_{j}{}^{m} \delta G^{(1)}_{lm} -  \frac{1}{2} \dB_{i}{}^{k} \dB^{ij} 
\dG_{j}{}^{l} \dG_{k}{}^{m} \delta G^{(1)}_{lm} -  \frac{1}{2} 
\dG_{i}{}^{k} \dG^{ij} \dG_{j}{}^{l} \dG_{k}{}^{m} \delta G^{(1)}_{lm}\nn\\&& -  
\frac{3}{8} \dB_{i}{}^{k} \dB^{ij} \dB_{j}{}^{l} \dB_{kl} \delta 
n^{(1)} + \frac{3}{4} \dB^{ij} \dB^{kl} \dG_{ik} \dG_{jl} \delta n^{(1)} -  
\frac{3}{2} \dB_{i}{}^{k} \dB^{ij} \dG_{j}{}^{l} \dG_{kl} \delta 
n^{(1)}\nn\\&& -  \frac{3}{8} \dG_{i}{}^{k} \dG^{ij} \dG_{j}{}^{l} \dG_{kl} 
\delta n^{(1)} -  \frac{1}{8} \dB_{i}{}^{k} \dB^{ij} \dB_{j}{}^{l} 
\dB_{kl} \delta \phi^{(1)} + \frac{1}{4} \dB^{ij} \dB^{kl} \dG_{ik} 
\dG_{jl} \delta \phi^{(1)}\nn\\&& -  \frac{1}{2} \dB_{i}{}^{k} \dB^{ij} 
\dG_{j}{}^{l} \dG_{kl} \delta \phi^{(1)} -  \frac{1}{8} \dG_{i}{}^{k} 
\dG^{ij} \dG_{j}{}^{l} \dG_{kl} \delta \phi^{(1)}\Bigg]\,,\labell{d2}
\eeqa
where we have used the fact that the lapse function appears in the $O(d,d)$-invariant action \reef{action2} by replacing $dt$ with $dt/n^3$.

By inserting the first-order perturbations at order $\alpha'$ \reef{dG1dB1} into \reef{d1}, \reef{d2}, and inserting the arbitrary first-order perturbations at order $\alpha'^2$ into \reef{dS0c1}, one finds that the cosmological action \reef{S0122} can be written as:
\beqa
S_2^c=\!\!\bS_2^c+\delta_1S_1^c-\frac{1}{2}\delta_1^2S_0^c+\delta_2S_0^c+\cJ_2=0\,,\labell{action21}
\eeqa
for some specific values of the parameters in field redefinitions and total derivative terms. Since we are not interested in studying the couplings at order $\alpha'^3$, we don't write the explicit form of the first-order field redefinitions  at order $\alpha'^2$ here.

For the bosonic theory, the calculations are the same except that the cosmological reduction of \reef{finalB} is given by:
\beqa
\bS_2^c&\!\!\!\!\!=\!\!\!\!\!&-\frac{2a_1^2}{\kappa^2}\int dt e^{-\Phi}\Bigg[\frac{11}{12} \dB_{i}{}^{k} \dB^{ij} \dB_{j}{}^{l} \dB_{k}{}^{m} 
\dB_{l}{}^{n} \dB_{mn} + \frac{1}{4} \dB_{ij} \dB^{ij} 
\dB_{k}{}^{m} \dB^{kl} \dB_{l}{}^{n} \dB_{mn}\labell{cB}\\&& + \frac{1}{48} 
\dB_{ij} \dB^{ij} \dB_{kl} \dB^{kl} \dB_{mn} \dB^{mn} + \frac{3}{2} 
\dB_{i}{}^{k} \dB^{ij} \dB_{l}{}^{n} \dB^{lm} \dG_{jm} \dG_{kn} + 
\frac{3}{2} \dB_{i}{}^{k} \dB^{ij} \dB_{j}{}^{l} \dB^{mn} \dG_{km} 
\dG_{ln} \nn\\&&-  \frac{1}{4} \dB_{ij} \dB^{ij} \dB^{kl} \dB^{mn} 
\dG_{km} \dG_{ln} -  \frac{3}{4} \dB^{ij} \dB^{kl} \dG_{i}{}^{m} 
\dG_{j}{}^{n} \dG_{km} \dG_{ln} + \frac{1}{4} \dB_{i}{}^{k} 
\dB^{ij} \dB_{l}{}^{n} \dB^{lm} \dG_{jk} \dG_{mn}\nn\\&& -  \frac{1}{2} 
\dB_{i}{}^{k} \dB^{ij} \dB_{j}{}^{l} \dB_{k}{}^{m} \dG_{l}{}^{n} 
\dG_{mn} -  \frac{1}{4} \dB_{ij} \dB^{ij} \dB_{k}{}^{m} \dB^{kl} 
\dG_{l}{}^{n} \dG_{mn} + \dB^{ij} \dB^{kl} \dG_{ik} \dG_{j}{}^{m} 
\dG_{l}{}^{n} \dG_{mn} \nn\\&&-  \frac{1}{2} \dB_{i}{}^{k} \dB^{ij} 
\dG_{j}{}^{l} \dG_{k}{}^{m} \dG_{l}{}^{n} \dG_{mn} + \frac{1}{6} 
\dG_{i}{}^{k} \dG^{ij} \dG_{j}{}^{l} \dG_{k}{}^{m} \dG_{l}{}^{n} 
\dG_{mn} -  \frac{1}{16} \dB_{ij} \dB^{ij} \dG_{k}{}^{m} \dG^{kl} 
\dG_{l}{}^{n} \dG_{mn} \nn\\&&-  \frac{1}{16} \dG_{ij} \dG^{ij} 
\dG_{k}{}^{m} \dG^{kl} \dG_{l}{}^{n} \dG_{mn} + \frac{1}{4} 
\dB_{i}{}^{k} \dB^{ij} \dB_{j}{}^{l} \dB_{kl} \dG_{mn} \dG^{mn} -  
\frac{1}{8} \dB_{i}{}^{k} \dB^{ij} \dG_{j}{}^{l} \dG_{kl} \dG_{mn} 
\dG^{mn} \nn\\&&+ \frac{1}{48} \dG_{ij} \dG^{ij} \dG_{kl} \dG^{kl} 
\dG_{mn} \dG^{mn} + \dB^{ij} \dB^{kl} \ddot{B}_{ik} \ddot{B}_{jl} + 
\frac{1}{4} \dB^{ij} \dB^{kl} \ddot{B}_{ij} \ddot{B}_{kl} + \dB_{i}{}^{k} 
\dB^{ij} \ddot{B}_{j}{}^{l} \ddot{B}_{kl} \nn\\&&- 4 \dB_{i}{}^{k} \dB^{ij} \dB^{lm} 
\dG_{jl} \ddot{B}_{km} -  \frac{1}{2} \dB_{i}{}^{k} \dB^{ij} \dB^{lm} 
\dG_{jk} \ddot{B}_{lm} + \dB_{i}{}^{k} \dB^{ij} \dB_{j}{}^{l} 
\dG_{k}{}^{m} \ddot{B}_{lm} -  \frac{1}{2} \dB_{ij} \dB^{ij} \dB^{kl} 
\dG_{k}{}^{m} \ddot{B}_{lm}\nn\\&& -  \ddot{G}_{i}{}^{k} \ddot{G}^{ij} \ddot{G}_{jk} + 2 
\dB^{ij} \dB^{kl} \dG_{i}{}^{m} \dG_{km} \ddot{G}_{jl} - 2 \dB^{ij} 
\dB^{kl} \ddot{G}_{ik} \ddot{G}_{jl} -  \frac{1}{4} \dG^{ij} \dG^{kl} 
\ddot{G}_{ik} \ddot{G}_{jl} + \frac{1}{4} \dG^{ij} \dG^{kl} \ddot{G}_{ij} \ddot{G}_{kl} 
\nn\\&&-  \frac{5}{2} \dB_{i}{}^{k} \dB^{ij} \ddot{G}_{j}{}^{l} \ddot{G}_{kl} + 
\frac{3}{2} \dG_{i}{}^{k} \dG^{ij} \ddot{G}_{j}{}^{l} \ddot{G}_{kl} + 
\frac{5}{2} \dB_{i}{}^{k} \dB^{ij} \dG_{j}{}^{l} \dG_{l}{}^{m} 
\ddot{G}_{km}-  
\frac{1}{4} \dG_{i}{}^{k} \dG^{ij} \dG_{jk} \dG^{lm} \ddot{G}_{lm} \nn\\&& - 2 
\dB^{ij} \dB^{kl} \dG_{ik} \dG_{j}{}^{m} \ddot{G}_{lm} -  \frac{1}{2} 
\dG_{i}{}^{k} \dG^{ij} \dG_{j}{}^{l} \dG_{k}{}^{m} \ddot{G}_{lm} + 
\frac{1}{4} \dB_{ij} \dB^{ij} \dG_{k}{}^{m} \dG^{kl} \ddot{G}_{lm} -  \frac{1}{4} \dB_{ij} \dB^{ij} \ddot{G}_{kl} \ddot{G}^{kl}\Bigg].\nn
\eeqa
Using the same steps as in the heterotic case, one finds:
\beqa
 S_2^c&=&\!\!\bS_2^c+\delta_1S_1^c-\frac{1}{2}\delta_1^2S_0^c+\delta_2S_0^c+\cJ_2=-\frac{2a_1^2}{\kappa^2}\int dt e^{-\Phi}\Bigg[\frac{1}{12} \dB_{i}{}^{k} \dB^{ij} \dB_{j}{}^{l} \dB_{k}{}^{m} 
\dB_{l}{}^{n} \dB_{mn} \nn\\&&+ \frac{1}{4} \dB_{i}{}^{k} \dB^{ij} 
\dB_{l}{}^{n} \dB^{lm} \dG_{jm} \dG_{kn} -  \frac{1}{2} 
\dB_{i}{}^{k} \dB^{ij} \dB_{j}{}^{l} \dB^{mn} \dG_{km} \dG_{ln} + \frac{1}{4} \dB^{ij} \dB^{kl} \dG_{i}{}^{m} \dG_{j}{}^{n} \dG_{km} 
\dG_{ln}\nn\\&& + \frac{1}{2} \dB_{i}{}^{k} \dB^{ij} \dB_{j}{}^{l} 
\dB_{k}{}^{m} \dG_{l}{}^{n} \dG_{mn} -  \frac{1}{2} \dB^{ij} 
\dB^{kl} \dG_{ik} \dG_{j}{}^{m} \dG_{l}{}^{n} \dG_{mn} + \frac{1}{2} 
\dB_{i}{}^{k} \dB^{ij} \dG_{j}{}^{l} \dG_{k}{}^{m} \dG_{l}{}^{n} 
\dG_{mn} \nn\\&&+ \frac{1}{12} \dG_{i}{}^{k} \dG^{ij} \dG_{j}{}^{l} 
\dG_{k}{}^{m} \dG_{l}{}^{n} \dG_{mn}\Bigg],\nn\\&&
  =-\frac{2}{\kappa^2}\left(\frac{-1}{3\times 2^6}\right)\int dt e^{-\phi}\tr(\dS^6),\labell{action41}
 \eeqa
for some specific values of the parameters in field redefinitions and total derivative terms. It is consistent with the cosmological action given by Eq. (\ref{cosm}). The results in Eq. (\ref{action21}) and Eq. (\ref{action41}) coincide with those found in \cite{Codina:2021cxh} through the study of the pure gravity parts of the couplings. This concludes our illustration that the couplings in Eq. (\ref{finalB}) in the bosonic string theory, as well as the couplings in Eq. (\ref{finalodd}) and Eq. (\ref{finaleven}) in the heterotic theory, are consistent with the cosmological symmetry.

\section{Consistency with S-matrix elements}

In this section, we compare the 4-point S-matrix elements constructed from the couplings presented in Section 1 with the low-energy expansion of the 4-point sphere-level S-matrix elements in string theory. 

It has been observed in \cite{Kawai:1985xq} that the sphere-level closed string S-matrix element can be written in terms of disk-level S-matrix elements. The disk-level S-matrix element of four gauge bosons' vertex operators on the boundary of the disk has been calculated in \cite{Schwarz:1982jn}. This amplitude for the $(s-t)$ channel is 
\beqa
\cA (\alpha' s,\alpha' t)&\sim&\frac{\Gamma(-\alpha's)\Gamma(-\alpha' t)}{\Gamma(1+\alpha' u)}K\,,
\eeqa
where the Mandelstam variables are:
\beqa
s=-(k_1+k_2)^2,\quad t=-(k_1+k_4)^2,\quad u=-(k_1+k_3)^2;\quad s+t+u=0\,.
\eeqa
The kinematic factor $K$ includes contractions of open string momenta $k_1,k_2,k_3,k_4$ with themselves and with the polarizations $\z_1,\z_2,\z_3,\z_4$. We are interested only in the terms in $K$ where the momenta are contracted with themselves. The kinematic factor for these terms in superstring and bosonic string theories is
\beqa
K^{S}(\alpha')&=&\alpha'^2st\z_1\cdot\z_3\,\z_2\cdot\z_4+\alpha'^2su\z_2\cdot\z_3\,\z_1\cdot\z_4+\alpha'^2tu\z_1\cdot\z_2\,\z_3\cdot\z_4+\cdots,\\
K^{B}(\alpha')&=&\left(\frac{\alpha'^2st}{1+\alpha'u}\right)\z_1\cdot\z_3\,\z_2\cdot\z_4+\left(\frac{\alpha'^2su}{1+\alpha't}\right)\z_2\cdot\z_3\,\z_1\cdot\z_4+\left(\frac{\alpha'^2tu}{1+\alpha's}\right)\z_1\cdot\z_2\,\z_3\cdot\z_4+\cdots,\nn
\eeqa
where dots represent the terms where the momenta and polarizations are contracted, which we are not interested in. Note that the above kinematic factors are $stu$-symmetric.
If the above amplitude is for right-moving modes, then there are similar amplitudes $\bar{\cA}$ for the left-moving modes with momenta $k_1,k_2,k_3,k_4$ and polarizations $\bar{\z_1},\bar{\z_2},\bar{\z_3},\bar{\z_4}$.

The closed string amplitude of four massless NS-NS states with momenta $k_i$ and polarizations $\epsilon_i=\z_i\bar{\z_i}$ for $i=1,2,3,4$ is then given as \cite{Kawai:1985xq}:
\beqa
A&=&-\left(\frac{\kappa^2}{\pi\alpha'}\right)\sin(\frac{\alpha'\pi}{2}k_2\cdot k_3)\cA(\frac{\alpha'}{4}s,\frac{\alpha'}{4}t)\bar{\cA}(\frac{\alpha'}{4}t,\frac{\alpha'}{4}u),\nn\\
&=&-\left(\frac{\kappa^2}{\alpha'}\right)\frac{\Gamma(-\frac{\alpha'}{4}s)\Gamma(-\frac{\alpha'}{4}t)\Gamma(-\frac{\alpha'}{4}u)}{\Gamma(1+\frac{\alpha'}{4}s)\Gamma(1+\frac{\alpha'}{4}t)\Gamma(1+\frac{\alpha'}{4}u)}K(\frac{\alpha'}{4})\bar{K}(\frac{\alpha'}{4}),
\eeqa
where we have normalized the amplitude to be consistent with the leading-order action \reef{action1}. The amplitude for the bosonic theory is given by replacing $K\bar{K}$ with $K^B\bar{K}^B$, and for the heterotic theory, one has to replace it with $K^B\bar{K}^S$. 

To find the $\alpha'$-expansion of the above closed S-matrix elements, one has to expand the Gamma functions and the tachyon poles in the bosonic kinematic factor $K^B$. After the $\alpha'$-expansion and replacing $\z_i\bar{\z_i}$ with the closed string polarization tensor $\epsilon_i$, one finds some terms where the momenta are contracted with the closed string polarization, which we are not interested in. In the terms where the momenta are contracted with themselves, there are still two sets of terms. One set includes two-trace terms, where each trace involves contractions of two polarizations. The other set involves one-trace terms, which involve contractions of all polarizations. We are going to compare the one-trace terms in the $\alpha'$-expansion of the closed string S-matrix element with the corresponding terms in the field theory using the actions presented in the Introduction section.

The one-trace terms in the heterotic closed string amplitude up to order $\alpha'^2$ when the polarization tensor is for a graviton or for the $B$-field are
\beqa
A^H&\!\!\!\!\!\!\!\!=\!\!\!\!\!\!\!\!\!& \frac{\kappa^2}{4}\Big[s ( 1-\frac{\alpha'}{4} t + \frac{\alpha'^2}{16}t^2+\cdots)\Tr(\epsilon_1\epsilon_3^T\epsilon_2\epsilon_4^T) +u (1 - 
\frac{\alpha'}{4} t +\frac{\alpha'^2}{16} t^2+\cdots) \Tr(\epsilon_1\epsilon_2^T\epsilon_3\epsilon_4^T)\\&&+
s (1 -\frac{\alpha'}{4} u +\frac{\alpha'^2}{16}u^2+\cdots)  \Tr(\epsilon_1\epsilon_4^T\epsilon_2\epsilon_3^T)+t (1 - \frac{\alpha'}{4}u + \frac{\alpha'^2}{16}u^2+\cdots) 
\Tr(\epsilon_1\epsilon_2^T\epsilon_4\epsilon_3^T)\nn\\&& +u(1 -\frac{\alpha'}{4} s + \frac{\alpha'^2}{16}s^2+\cdots)  
\Tr(\epsilon_1\epsilon_4^T\epsilon_3\epsilon_2^T) +t (1 - \frac{\alpha'}{4} s +\frac{\alpha'^2}{16} s^2+\cdots)
\Tr(\epsilon_1\epsilon_3^T\epsilon_4\epsilon_2^T)+\cdots\Big].\nn
\eeqa
where dots in each parenthesis represent the higher orders of $\alpha'$, and the dots at the end of the equation represent terms in which momenta contract with the polarizations and terms that involve the two-traces. The corresponding terms in the bosonic closed string amplitude are
\beqa
A^B&=&   
\frac{\kappa^2s}{4} \bigl(1 +\frac{\alpha'}{4} s + \frac{\alpha'^2}{16}s^2 - \frac{\alpha'^2}{16}tu+\cdots\bigr) \Big[\Tr(\epsilon_1\epsilon_3^T\epsilon_2\epsilon_4^T)+\Tr(\epsilon_1\epsilon_4^T\epsilon_2\epsilon_3^T)\Big] \nn\\&&+ 
\frac{\kappa^2u}{4} \bigl(1 +\frac{\alpha'}{4} u + \frac{\alpha'^2}{16}u^2 - \frac{\alpha'^2}{16}st+\cdots\bigr) \Big[
 \Tr(\epsilon_1\epsilon_2^T\epsilon_3\epsilon_4^T)+\Tr(\epsilon_1\epsilon_4^T\epsilon_3\epsilon_2^T)\Big]\nn\\&& +\frac{\kappa^2t}{4} \bigl(1 +\frac{\alpha'}{4} t + \frac{\alpha'^2}{16}t^2 - \frac{\alpha'^2}{16}su+\cdots\bigr) \Big[\Tr(\epsilon_1\epsilon_2^T\epsilon_4\epsilon_3^T)+\Tr(\epsilon_1\epsilon_3^T\epsilon_4\epsilon_2^T)\Big] +\cdots.
\eeqa
Using the identity $\Tr(ABCD)=\Tr(A^TD^TC^TB^T)$, one can write both amplitudes in terms of $\Tr(\epsilon_1\epsilon_3\epsilon_2\epsilon_4)$, $\Tr(\epsilon_1\epsilon_2\epsilon_3\epsilon_4)$, and $\Tr(\epsilon_1\epsilon_2\epsilon_4\epsilon_3)$. 

When the four states are gravitons or they are $B$-fields, one can write the amplitudes as
\beqa
A&=&f(s,t,u)\Tr(\epsilon_1\epsilon_3\epsilon_2\epsilon_4) +f(u,t,s) \Tr(\epsilon_1\epsilon_2\epsilon_3\epsilon_4)+f(t,u,s)
\Tr(\epsilon_1\epsilon_2\epsilon_4\epsilon_3)+\cdots\,,\labell{Agggg}
\eeqa
where the function $f(s,t,u)$ is
\beqa
f^H(s,t,u)&=& \frac{\kappa^2}{2} \Big[s+\frac{\alpha'}{8}s^2 + \frac{\alpha'^2}{32}(s^3-2stu)+\cdots\Big],\nn\\
f^B(s,t,u)&=& \frac{\kappa^2}{2} \Big[s+\frac{\alpha'}{4}s^2 + \frac{\alpha'^2}{16}(s^3-stu)+\cdots\Big].\labell{fHB}
\eeqa
When states 1 and 2 are gravitons and states 3 and 4 are $B$-fields, the amplitudes can be written as:
\beqa
A&=&f(s,t,u)\Tr(\epsilon_1\epsilon_3\epsilon_2\epsilon_4) -f(u,t,s) \Tr(\epsilon_1\epsilon_2\epsilon_3\epsilon_4)-f(t,u,s)
\Tr(\epsilon_1\epsilon_2\epsilon_4\epsilon_3)+\cdots\,,\labell{AggBB}
\eeqa
where $f$ is  the same function as in \reef{fHB}.  

In the heterotic theory, when state 1 is a graviton and states 2, 3, and 4 are $B$-fields, or when state 4 is a $B$-field and states 1, 2, and 3 are gravitons, the amplitude can be written as:
\beqa
A^H&=&g(s,t,u)\Tr(\epsilon_1\epsilon_3\epsilon_2\epsilon_4) +g(u,t,s) \Tr(\epsilon_1\epsilon_2\epsilon_3\epsilon_4)+g(t,u,s)
\Tr(\epsilon_1\epsilon_2\epsilon_4\epsilon_3)+\cdots\,,\labell{ABBBg}
\eeqa
where the function $g(s,t,u)$ is
\beqa
g(s,t,u)&=& \kappa^2(t-u)\Big [-\frac{\alpha'}{16}s + \frac{\alpha'^2}{64}s^2+\cdots\Big]\,.\labell{gH}
\eeqa
It has been shown in \cite{Metsaev:1986yb} that when all four states are gravitons, the amplitude \reef{Agggg} for both the bosonic and heterotic theories is reproduced by the couplings in \reef{action1}, \reef{Mis}, \reef{finalB}, and \reef{CS}. In the next subsection, we will demonstrate that the string amplitudes of all other states at order $\alpha'^2$ are exactly reproduced by the couplings in \reef{action1}, \reef{Mis}, \reef{finalB}, \reef{CS}, \reef{Misodd}, \reef{finalodd}, and \reef{finaleven}.

\subsection{S-matrix elements in field theory}

In this subsection, we calculate the S-matrix elements in field theory. Since the couplings in the heterotic theory involve the Chern-Simons three-form $\Omega$, which is expressed in terms of frame fields $e_\mu{}^{\mu_1}$ and their partial derivatives, and the closed string theory amplitudes resulting from the KLT prescription involve the polarization of the graviton $\epsilon_{\mu\nu}$, we need to express $\Omega$ in terms of the metric and its partial derivatives. The metric can then be perturbed as $G_{\mu\nu}=\eta_{\mu\nu}+\kappa h_{\mu\nu}$. In calculating the S-matrix elements in field theory, we also need to write the fluctuation of the $B$-field as $\kappa b$.

By taking partial derivatives of the relation $e_\mu{}^{\mu_1}e_\nu{}^{\nu_1}\eta_{\mu_1\nu_1}=G_{\mu\nu}$, we can determine the following:
\beqa
\prt_\alpha e_\beta{}^{\mu_1}e_{\gamma\mu_1}+\prt_\alpha e_\gamma{}^{\mu_1}e_{\beta\mu_1}&=&\prt_\alpha G_{\beta\gamma}\,.
\eeqa
Then one can assume that the two terms on the left-hand side above are identical. In this case, we have the relation:
\beqa
\partial_\alpha e_\beta{}^{\mu_1}e_{\gamma\mu_1}&=&\frac{1}{2}\partial_\alpha G_{\beta\gamma}\,.
\eeqa
Using this relation, we can express $\Omega$ as:
\beqa
\Omega_{\alpha\beta\gamma}&=&G^{\kappa[\delta}G^{\epsilon]\eta}\prt_{\ka}G_{[\alpha\eta}\prt_{\gamma}
\prt_{\delta}G_{\beta]\epsilon}+G^{\delta\theta}G^{\epsilon\tau}G^{\mu\kappa}\Big[\frac{1}{2}\prt_\delta G_{[\gamma\mu}\prt_{\epsilon}G_{\beta\kappa}\prt_\theta G_{\alpha]\tau}\nn\\&&
\qquad\qquad\qquad\qquad\qquad+\frac{1}{4}\prt_{[\alpha}G_{\theta\tau}\prt_{\gamma}G_{\delta\kappa}\prt_{\epsilon}G_{\beta]\mu}+\frac{1}{6}\prt_{\epsilon}G_{[\gamma\mu}\prt_{\theta}G_{\alpha\tau}\prt_{\kappa}G_{\beta]\delta}\Big]\,.\labell{3form}
\eeqa
where the subscript antisymmetrization of indices is between $\alpha$, $\beta$, and $\gamma$. Note that $\Omega$ has at least two gravitons.

The four-point function has both contact terms and massless pole contributions in which the graviton and/or $B$-field propagate between two vertices in the latter cases. Using the leading-order action (\ref{action1}), one finds the following graviton and $B$-field propagators:
\beqa
(\tG_h)_{\mu\nu, \lambda\rho}&=&\frac{1}{2k^2}\left(\eta_{\mu\lambda}\eta_{\nu\rho}+\eta_{\mu\rho}\eta_{\nu\lambda}-\frac{1}{\frac{D}{2}-1}\eta_{\mu\nu}\eta_{\lambda\rho}\right),\nn\\
(\tG_b)_{\mu\nu, \lambda\rho}&=&\frac{1}{2k^2}\left(\eta_{\mu\lambda}\eta_{\nu\rho}-\eta_{\mu\rho}\eta_{\nu\lambda}\right),
\eeqa
where $k^\mu$ is the momentum of the graviton or $B$-field in the propagator, and we have used a Euclidean signature spacetime metric in which the partition function is $Z\sim\int e^{-S}$. 

\subsubsection{Four $B$-fields amplitude}

The contact term of four $B$-fields at order $\alpha'^2$ in the heterotic action (\ref{finaleven}) produces the following one-trace terms in the field theory amplitude:
\beqa
A^H_{contact}&=&\frac{\kappa^2\alpha'^2}{128}t (s^2 + s t + t^2)\Tr(\epsilon_1\epsilon_3\epsilon_4\epsilon_2) +
 \frac{\kappa^2\alpha'^2}{128} s (s^2 + s t + t^2)\Tr(\epsilon_1\epsilon_4\epsilon_2\epsilon_3 )\nn\\&&- 
 \frac{\kappa^2\alpha'^2}{128} (s^3 + 2 s^2 t + 2 s t^2 + 
    t^3)\Tr(\epsilon_1\epsilon_4\epsilon_3\epsilon_2)+\cdots\,,
\eeqa
where dots represent the two-trace terms and terms in which momenta are contracted with the polarization tensors that we are not interested in.

The pole amplitude resulting from two vertices at order $\alpha'$ and one propagator in between produces the following one-trace terms in the amplitude:
\beqa
A^H_{pole}&=&\tV_4(bbh)\tG_h\tV_4(hbb)=\frac{\kappa^2\alpha'^2}{128} (-s^3 - u^3) \Tr(\epsilon_1\epsilon_3\epsilon_4\epsilon_2)\\&&+
 \frac{\kappa^2\alpha'^2}{128} s (s^2 + 3 s u + 3 u^2)\Tr(\epsilon_1\epsilon_4\epsilon_2\epsilon_3 )+ 
 \frac{\kappa^2\alpha'^2}{128}u (3 s^2 + 3 s u + u^2)\Tr(\epsilon_1\epsilon_4\epsilon_3\epsilon_2)+\cdots.\nn
\eeqa
Note that there is no pole in the one-trace terms. The subscript 4 in the vertex means it has four momenta. Additionally, there are no couplings between two $B$-fields and one graviton in the action (\ref{finaleven}) at the six-derivative order. Hence, there is no amplitude with the structure $\tV_2(bbh)\tG_h\tV_6(hbb)$.
By summing the terms $A^H_{contact}$ and $A^H_{pole}$, one finds that they are exactly the same as the corresponding amplitude in (\ref{Agggg}).

The contact term of four $B$-fields at order $\alpha'^2$ in the bosonic action (\ref{finalB}) produces the following one-trace terms in the field theory amplitude:
\beqa
A^B_{contact}&=&\frac{\kappa^2\alpha'^2}{16}stu\Big[\Tr(\epsilon_1\epsilon_3\epsilon_4\epsilon_2) +
\Tr(\epsilon_1\epsilon_4\epsilon_2\epsilon_3 )+
\Tr(\epsilon_1\epsilon_4\epsilon_3\epsilon_2)\Big]+\cdots\,.
\eeqa
The pole amplitude resulting from two vertices at order $\alpha'$ and one propagator in between produces the following one-trace terms in the amplitude:
\beqa
A^B_{pole}&=&\tV_4(bbh)\tG_h\tV_4(hbb)=\frac{\kappa^2\alpha'^2}{32} (-s^3 - u^3) \Tr(\epsilon_1\epsilon_3\epsilon_4\epsilon_2)\\&&+
 \frac{\kappa^2\alpha'^2}{32} s (s^2 + 3 s u + 3 u^2)\Tr(\epsilon_1\epsilon_4\epsilon_2\epsilon_3 )+ 
 \frac{\kappa^2\alpha'^2}{32}u (3 s^2 + 3 s u + u^2)\Tr(\epsilon_1\epsilon_4\epsilon_3\epsilon_2)+\cdots.\nn
\eeqa
Here, also, there are no couplings between two $B$-fields and one graviton in the action (\ref{finalB}) at the six-derivative order. Hence, there is no amplitude with the structure $\tV_2(bbh)\tG_h\tV_6(hbb)$.
By summing the terms $A^B_{contact}$ and $A^B_{pole}$, one finds that they are exactly the same as the corresponding amplitude in (\ref{Agggg}).

\subsubsection{ Two $B$-fields amplitude}

The contact terms of two $B$-fields and two gravitons at order $\alpha'^2$ in the heterotic action (\ref{finaleven}) produce the following one-trace terms in the field theory amplitude:
\beqa
A^H_{contact}&=&\frac{\kappa^2\alpha'^2}{128} s (s^2 + 2 s t + 2 t^2)\Tr(\epsilon_1\epsilon_3\epsilon_4\epsilon_2) +
 \frac{\kappa^2\alpha'^2}{64} s (s^2 + 2 s t + 2 t^2)\Tr(\epsilon_1\epsilon_4\epsilon_2\epsilon_3 )\nn\\&&+
 \frac{\kappa^2\alpha'^2}{128}s (s^2 + 2 s t + 2 t^2)\Tr(\epsilon_1\epsilon_4\epsilon_3\epsilon_2)+\cdots\,.
\eeqa
The pole amplitude resulting from two vertices at order $\alpha'$ and one propagator in between produces the following one-trace terms in the amplitudes:
\beqa
A_1^H{}_{pole}&=&\tV_4(bbh)\tG_h\tV_4(hhh)=\frac{\kappa^2\alpha'^2}{128} s^3 \Big[ \Tr(\epsilon_1\epsilon_2\epsilon_3\epsilon_4)+
\Tr(\epsilon_1\epsilon_2\epsilon_4\epsilon_3 )\Big]+\cdots,\nn\\
A_2^H{}_{pole}&=&\tV_4(hbh)\tG_h\tV_4(hbh)=\frac{\kappa^2\alpha'^2}{128} t^3 \Big[ \Tr(\epsilon_1\epsilon_2\epsilon_3\epsilon_4)-
\Tr(\epsilon_1\epsilon_4\epsilon_2\epsilon_3 )\Big]\nn\\&&\qquad\qquad\qquad\qquad\quad+\frac{\kappa^2\alpha'^2}{128} u^3 \Big[ \Tr(\epsilon_1\epsilon_2\epsilon_4\epsilon_3)-
\Tr(\epsilon_1\epsilon_4\epsilon_2\epsilon_3 )\Big]\cdots,\nn\\
A_3^H{}_{pole}&=&\tV_4(hbb)\tG_b\tV_4(bbh)=\frac{\kappa^2\alpha'^2}{128} t^3 \Big[ \Tr(\epsilon_1\epsilon_2\epsilon_3\epsilon_4)+
\Tr(\epsilon_1\epsilon_4\epsilon_2\epsilon_3 )\Big]\nn\\&&\qquad\qquad\qquad\qquad\quad+\frac{\kappa^2\alpha'^2}{128} u^3 \Big[ \Tr(\epsilon_1\epsilon_2\epsilon_4\epsilon_3)+
\Tr(\epsilon_1\epsilon_4\epsilon_2\epsilon_3 )\Big]\cdots.
\eeqa
Note that since the Chern-Simons three-form in (\ref{3form}) has at least two gravitons, the pure gravity couplings at order $\alpha'^2$ in (\ref{CS}) have no three-graviton vertex, i.e., $\tV_6(hhh)=0$. Moreover, we have used the fact that $\tV_2(hbh)=0$.
By summing the terms $A^H_{contact}$, $A_1^H{}_{pole}$, $A_2^H{}_{pole}$, and $A_3^H{}_{pole}$, one finds that they are exactly the same as the corresponding amplitude in (\ref{AggBB}).

The contact terms of two $B$-fields and two gravitons at order $\alpha'^2$ in the bosonic action (\ref{finalB}) produce the following one-trace terms in the field theory amplitude:
\beqa
A^B_{contact}&=&\frac{\kappa^2\alpha'^2}{64}  s (3 s^2 + 4 s t + 4 t^2)\Tr(\epsilon_1\epsilon_3\epsilon_4\epsilon_2) +
 \frac{\kappa^2\alpha'^2}{16} s (s^2 + 2 s t + 2 t^2)\Tr(\epsilon_1\epsilon_4\epsilon_2\epsilon_3 )\nn\\&&+
 \frac{\kappa^2\alpha'^2}{64}s (3 s^2 + 4 s t + 4 t^2)\Tr(\epsilon_1\epsilon_4\epsilon_3\epsilon_2)+\cdots\,.
\eeqa
The pole amplitude resulting from two vertices and one propagator in between produces the following one-trace terms in the amplitudes:
\beqa
A_1^B{}_{pole}&=&\tV_4(bbh)\tG_h\tV_4(hhh)=\frac{\kappa^2\alpha'^2}{32} s^3 \Big[ \Tr(\epsilon_1\epsilon_2\epsilon_3\epsilon_4)+
\Tr(\epsilon_1\epsilon_2\epsilon_4\epsilon_3 )\Big]+\cdots,\nn\\
A_2^B{}_{pole}&=&\tV_4(bhb)\tG_b\tV_4(bhb)=-\frac{\kappa^2\alpha'^2}{32} s (s^2 + 3 s u + 3 u^2)  
\Tr(\epsilon_1\epsilon_4\epsilon_2\epsilon_3 )\nn\\&&\qquad\qquad\qquad\qquad\quad+\frac{\kappa^2\alpha'^2}{32} u^3 \Tr(\epsilon_1\epsilon_3\epsilon_4\epsilon_2)+
\frac{\kappa^2\alpha'^2}{32} t^3\Tr(\epsilon_1\epsilon_4\epsilon_3\epsilon_2 )\cdots,\nn\\
A_3^B{}_{pole}&=&\tV_2(bbh)\tG_h\tV_6(hhh)=-\frac{3\kappa^2\alpha'^2}{64} s^3 \Big[ \Tr(\epsilon_1\epsilon_2\epsilon_3\epsilon_4)+
\Tr(\epsilon_1\epsilon_2\epsilon_4\epsilon_3 )\Big]+\cdots.
\eeqa
Note that in the bosonic action at order $\alpha'^2$ in (\ref{finalB}), there is a non-zero three-graviton vertex which has been used in the last relation above. 
By summing the terms $A^B_{contact}$, $A_1^B{}_{pole}$, $A_2^B{}_{pole}$, and $A_3^B{}_{pole}$, one finds that they are exactly the same as the corresponding amplitude in (\ref{AggBB}).

\subsubsection{Three $B$-fields amplitude}

The contact terms of three $B$-fields and one graviton at order $\alpha'^2$ in the heterotic action (\ref{finalodd}) produce the following one-trace terms in the field theory amplitude:
\beqa
A^H_{contact}&=&-\frac{\kappa^2\alpha'^2}{128}  (s - u) (t^2 + s u)\Tr(\epsilon_1\epsilon_3\epsilon_4\epsilon_2) +
 \frac{\kappa^2\alpha'^2}{128}(-t + u) (s^2 + t u)\Tr(\epsilon_1\epsilon_4\epsilon_2\epsilon_3 )\nn\\&&+
 \frac{\kappa^2\alpha'^2}{128}(s - t) (s t + u^2)\Tr(\epsilon_1\epsilon_4\epsilon_3\epsilon_2)+\cdots\,.
\eeqa
The pole amplitude resulting from two vertices at order $\alpha'$ and one propagator in between produces the following one-trace terms in the amplitude:
\beqa
A^H_{pole}&=&\tV_4(bbh)\tG_h\tV_4(hbh)=-\frac{\kappa^2\alpha'^2}{128} (s^3 - u^3) \Tr(\epsilon_1\epsilon_3\epsilon_4\epsilon_2)+
\frac{\kappa^2\alpha'^2}{128}(-t^3 + u^3) \Tr(\epsilon_1\epsilon_4\epsilon_2\epsilon_3)\nn\\&&\qquad\qquad\qquad\qquad\quad+
\frac{\kappa^2\alpha'^2}{128}(s^3 - t^3) \Tr(\epsilon_1\epsilon_4\epsilon_3\epsilon_2)+\cdots\,.
\eeqa
One finds that the sum of $A^H_{contact}$ and $A^H_{pole}$ is exactly the same as the corresponding amplitude in (\ref{ABBBg}). 

\subsubsection{One $B$-field amplitude}

The odd-parity effective action (\ref{finalodd}) contains no coupling involving a single $B$-field. However, the odd-parity couplings in (\ref{Misodd}), resulting from replacing the non-standard $B$-field strength in the Meissner action, do involve a single $B$-field.
The contact term of three gravitons and one $B$-field at order $\alpha'^2$ in the heterotic action (\ref{Misodd}) produces the following one-trace terms in the field theory amplitude:
\beqa
A^H_{contact}&=&-\frac{\kappa^2\alpha'^2}{128}  (s - u) (t^2 + s u)\Tr(\epsilon_1\epsilon_3\epsilon_4\epsilon_2) - \frac{\kappa^2\alpha'^2}{128}(s-t) (u^2 + st)\Tr(\epsilon_1\epsilon_2\epsilon_4\epsilon_4 )\nn\\&&+
 \frac{\kappa^2\alpha'^2}{128}(t-u) (s^2+tu)\Tr(\epsilon_1\epsilon_3\epsilon_2\epsilon_4)+\cdots\,.
\eeqa
The pole amplitude resulting from two vertices at order $\alpha'$ and one propagator in between produces the following one-trace terms in the amplitude:
\beqa
A^H_{pole}&=&\tV_4(hhh)\tG_h\tV_4(hhb)=-\frac{\kappa^2\alpha'^2}{128} (s^3 - u^3) \Tr(\epsilon_1\epsilon_3\epsilon_4\epsilon_2)-
\frac{\kappa^2\alpha'^2}{128}(s^3-t^3) \Tr(\epsilon_1\epsilon_2\epsilon_3\epsilon_4)\nn\\&&\qquad\qquad\qquad\qquad\quad-
\frac{\kappa^2\alpha'^2}{128}(u^3-t^3) \Tr(\epsilon_1\epsilon_3\epsilon_2\epsilon_4)+\cdots\,.
\eeqa
One finds that the sum of $A^H_{contact}$ and $A^H_{pole}$ is exactly the same as the corresponding amplitude in (\ref{ABBBg}). This confirms our observation that the 4-point functions in the effective actions of the bosonic and heterotic theories at order $\alpha'^2$ are consistent with the corresponding 4-point sphere-level S-matrix elements at order $\alpha'^2$.

\section{Conclusion}

In this paper, using appropriate field redefinitions, we have expressed the even-parity couplings in the effective actions of both the bosonic string theory and the heterotic string theory at order $\alpha'^2$ in a canonical form where the dilaton appears only as the overall factor. These couplings, which have recently been discovered through T-duality, can be represented as \reef{finalB} and \reef{finaleven}, respectively. Additionally, we demonstrate that the cosmological reduction of the couplings in the bosonic theory, as well as the even- and odd-parity couplings in the heterotic theory, satisfy the $O(d,d)$ symmetry in the proposed canonical form put forward in \cite{Hohm:2015doa, Hohm:2019jgu}. This achievement is accomplished by incorporating appropriate one-dimensional total derivative terms and utilizing suitable one-dimensional field redefinitions.

Since the original couplings were derived by imposing the T-duality symmetry $O(1,1)$ on the most general covariant couplings \cite{Garousi:2019mca, Garousi:2023kxw}, it is expected that the cosmological couplings remain invariant under $O(d,d)$ transformations. To further confirm the couplings, we extensively examine them by comparing the 4-point S-matrix elements in the effective actions at order $\alpha'^2$ with the corresponding sphere-level S-matrix elements in string theory. Remarkably, we establish an exact agreement between these two S-matrix elements.

The aforementioned S-matrix calculations provide confirmation of the couplings in field theory at order $\alpha'^2$ that involve two, three, and four $B$-fields. Additionally, the couplings described in \reef{finalB} and \reef{finaleven} involve six $B$-fields as well.
To validate these couplings through S-matrix elements, it is necessary to determine the 6-point sphere-level S-matrix elements of NS-NS vertex operators in both the bosonic and heterotic theories, and then expand them to isolate the terms at order $\alpha'^2$. By employing the KLT prescription \cite{Kawai:1985xq}, one needs to calculate the 6-point disk-level S-matrix element of gauge bosons in the bosonic and superstring theories, and subsequently obtain its $\alpha'$-expansion. Such calculations have been conducted in \cite{Oprisa:2005wu} for superstring theory. Therefore, it is necessary to find the 6-point disk-level S-matrix element in the bosonic theory and determine its $\alpha'$-expansion. Utilizing the KLT prescription, one can then deduce the NS-NS S-matrix elements at order $\alpha'^2$. These calculated S-matrix elements should align with the corresponding S-matrix elements in the effective actions at order $\alpha'^2$. The detailed calculations for this procedure are left for future work.

%\newpage

\end{document}